\documentclass[aps,prc, amsmath, amssymb,prc,twocolumn,letterpaper,superscriptaddress,floatfix]{revtex4-1}
\usepackage{graphicx,xcolor,textpos,tabularx}
\usepackage{epstopdf}
\usepackage{amsmath}
\usepackage{xcolor}
\usepackage{import}

\usepackage{graphicx}
\usepackage{dcolumn}
\usepackage{bm}
\usepackage{hyperref}

\usepackage{titlesec}

\titleformat{\paragraph}
{\normalfont\normalsize\bfseries}{\theparagraph}{1em}{}
\titlespacing*{\paragraph}
{0pt}{3.25ex plus 1ex minus .2ex}{1.5ex plus .2ex}

\begin{document}
\graphicspath{{./Figuren/}}

\title{Shape staggering of mid-shell mercury isotopes from in-source laser spectroscopy compared with 
Density Functional Theory and Monte Carlo Shell Model calculations}

\author{S.~Sels}
\email{Simon.Sels@kuleuven.be}
\email{Simon.Sels@cern.ch}
\altaffiliation[\\ Current address: ]{CERN, CH-1211 Geneve 23, Switzerland}
\affiliation{KU Leuven, Instituut voor Kern- en Stralingsfyisca, B-3001 Leuven, Belgium}

\author{T.~Day Goodacre}%
\affiliation{School of Physics and Astronomy, The University of Manchester, Manchester M13 9PL, UK}
\affiliation{CERN, CH-1211 Geneve 23, Switzerland}

\author{B. A.~Marsh}%
\affiliation{CERN, CH-1211 Geneve 23, Switzerland}

\author{A.~Pastore}%
\affiliation{Department of Physics, University of York, York Y010 5DD, UK}

\author{W.~Ryssens}
\affiliation{IPNL, Universit\'e de Lyon, Universit\'e Lyon 1, CNRS/IN2P3, F-69622 Villeurbanne, France}

\author{Y. ~Tsunoda}
\affiliation{Department of Physics, The University of Tokyo, Bunkyo-ku, 113-0033 Tokyo, Japan}

\author{N.~Althubiti}%
\affiliation{School of Physics and Astronomy, The University of Manchester, Manchester M13 9PL, UK}

\author{B.~Andel}%
\affiliation{Department of Nuclear Physics and Biophysics, Comenius University in Bratislava, 84248 Bratislava, Slovakia}

\author{A. N.~Andreyev}%
\affiliation{Department of Physics, University of York, York Y010 5DD, UK}
\affiliation{Advanced Science Research Center, Japan Atomic Energy Agency (JAEA), Tokai-mura, Japan}

\author{D.~Atanasov}%
\affiliation{Max-Planck-Institut f\"ur Kernphysik, Saupfercheckweg 1, 69117 Heidelberg, Germany}

\author{A. E.~Barzakh}%
\affiliation{Petersburg Nuclear Physics Institute, NRC Kurchatov Institute, Gatchina 188300, Russia}

\author{M.~Bender}
\affiliation{IPNL, Universit\'e de Lyon, Universit\'e Lyon 1, CNRS/IN2P3, F-69622 Villeurbanne, France}

\author{J.~Billowes}%
\affiliation{School of Physics and Astronomy, The University of Manchester, Manchester M13 9PL, UK}

\author{K.~Blaum}%
\affiliation{Max-Planck-Institut f\"ur Kernphysik, Saupfercheckweg 1, 69117 Heidelberg, Germany}

\author{T. E.~Cocolios}%
\affiliation{KU Leuven, Instituut voor Kern- en Stralingsfyisca, B-3001 Leuven, Belgium}

\author{J. G.~Cubiss}%
\affiliation{Department of Physics, University of York, York Y010 5DD, UK}

\author{J.~Dobaczewski}%
\affiliation{Department of Physics, University of York, York Y010 5DD, UK}
\affiliation{Institute of Theoretical Physics, Faculty of Physics, University of Warsaw, ul. Pasteura 5, PL-02-093 Warsaw, Poland}

\author{G. J.~Farooq-Smith}%
\affiliation{KU Leuven, Instituut voor Kern- en Stralingsfyisca, B-3001 Leuven, Belgium}

\author{D. V.~Fedorov}%
\affiliation{Petersburg Nuclear Physics Institute, NRC Kurchatov Institute, Gatchina 188300, Russia}

\author{V. N.~Fedosseev}%
\affiliation{CERN, CH-1211 Geneve 23, Switzerland}

\author{K. T.~Flanagan}%
\affiliation{School of Physics and Astronomy, The University of Manchester, Manchester M13 9PL, UK}

\author{L. P.~Gaffney}%
\affiliation{School of Engineering and Computing, University of the West of Scotland, Paisley, PA1 2BE, UK}
\affiliation{KU Leuven, Instituut voor Kern- en Stralingsfyisca, B-3001 Leuven, Belgium}

\author{L.~Ghys}%
\affiliation{Belgian Nuclear Research Center SCK$\bullet$CEN, Boeretang 200, B-2400 Mol, Belgium}
\affiliation{KU Leuven, Instituut voor Kern- en Stralingsfyisca, B-3001 Leuven, Belgium}

\author{P-H.~Heenen}
\affiliation{PNTPM, CP229, Universite Libre de Bruxelles, B-1050 Bruxelles, Belgium}

\author{M.~Huyse}%
\affiliation{KU Leuven, Instituut voor Kern- en Stralingsfyisca, B-3001 Leuven, Belgium}

\author{S.~Kreim}%
\affiliation{Max-Planck-Institut f\"ur Kernphysik, Saupfercheckweg 1, 69117 Heidelberg, Germany}

\author{D.~Lunney}%
\affiliation{CSNSM-IN2P3-CNRS, Universit\'e Paris-Sud, 91406 Orsay, France}

\author{K. M.~Lynch}%
\affiliation{CERN, CH-1211 Geneve 23, Switzerland}

\author{V.~Manea}%
\affiliation{Max-Planck-Institut f\"ur Kernphysik, Saupfercheckweg 1, 69117 Heidelberg, Germany}

\author{Y.~Martinez Palenzuela}%
\affiliation{KU Leuven, Instituut voor Kern- en Stralingsfyisca, B-3001 Leuven, Belgium}

\author{T. M.~Medonca}%
\affiliation{CERN, CH-1211 Geneve 23, Switzerland}

\author{P. L.~Molkanov}%
\affiliation{Petersburg Nuclear Physics Institute, NRC Kurchatov Institute, Gatchina 188300, Russia}


\author{T.~Otsuka}%
\affiliation{Department of Physics, The University of Tokyo, Bunkyo-ku, 113-0033 Tokyo, Japan}
\affiliation{National Superconducting Cyclotron Laboratory, Michigan State University, East Lansing, MI 48824, USA}
\affiliation{KU Leuven, Instituut voor Kern- en Stralingsfyisca, B-3001 Leuven, Belgium}



\author{J. P.~Ramos}%
\affiliation{CERN, CH-1211 Geneve 23, Switzerland}
\affiliation{Laboratory of Powder Technology, \'Ecole polytechnique f\'ed\'erale de Lausanne (EPFL), CH-1015 Lausanne, Switzerland}

\author{R. E.~Rossel}%
\affiliation{CERN, CH-1211 Geneve 23, Switzerland}
\affiliation{Insitut f\"urPhysik, Johannes Gutenberg-Universit\"at, 55122 Mainz, Germany}

\author{S.~Rothe}%
\affiliation{CERN, CH-1211 Geneve 23, Switzerland}

\author{L.~Schweikhard}%
\affiliation{Institut f\"ur Physik, Universit\"at Greifswald, 17487 Greifswald, Germany}

\author{M. D.~Seliverstov}%
\affiliation{Petersburg Nuclear Physics Institute, NRC Kurchatov Institute, Gatchina 188300, Russia}


\author{P.~Spagnoletti}%
\affiliation{School of Engineering and Computing, University of the West of Scotland, Paisley, PA1 2BE, UK}

\author{C.~Van Beveren}%
\affiliation{KU Leuven, Instituut voor Kern- en Stralingsfyisca, B-3001 Leuven, Belgium}

\author{P.~Van Duppen}%
\affiliation{KU Leuven, Instituut voor Kern- en Stralingsfyisca, B-3001 Leuven, Belgium}

\author{M.~Veinhard}%
\affiliation{CERN, CH-1211 Geneve 23, Switzerland}

\author{E.~Verstraelen}%
\affiliation{KU Leuven, Instituut voor Kern- en Stralingsfyisca, B-3001 Leuven, Belgium}

\author{A.~Welker}%
\affiliation{Technische Universitat Dresden, 01069 Dresden, Germany}

\author{K.~Wendt}%
\affiliation{Insitut f\"urPhysik, Johannes Gutenberg-Universit\"at, 55122 Mainz, Germany}

\author{F.~Wienholtz}%
\affiliation{Institut f\"ur Physik, Universit\"at Greifswald, 17487 Greifswald, Germany}

\author{R.N.~Wolf}%
\affiliation{Max-Planck-Institut f\"ur Kernphysik, Saupfercheckweg 1, 69117 Heidelberg, Germany}

\author{A.~Zadvornaya}%
\affiliation{KU Leuven, Instituut voor Kern- en Stralingsfyisca, B-3001 Leuven, Belgium}


\date{\today}

\begin{abstract}

Neutron-deficient $^{177-185}$Hg isotopes were studied using in-source laser resonance-ionization spectroscopy 
at the CERN-ISOLDE radioactive ion-beam facility, in an experiment combining different detection methods tailored to the studied isotopes. 
These include either $\alpha$-decay tagging or Multi-reflection Time-of-Flight gating to identify the isotopes of interest. 
The endpoint of the odd-even nuclear shape staggering in mercury was observed directly by measuring 
for the first time the isotope shifts and hyperfine structures of $^{177-180}$Hg. 
Changes in the mean-square charge radii for all mentioned isotopes, 
magnetic dipole and electric quadrupole moments of the odd-$A$ isotopes and arguments 
in favor of $I=7/2$ spin assignment for $^{177,179}$Hg were deduced. 
Experimental results are compared with Density Functional Theory (DFT) and Monte-Carlo Shell Model (MCSM) calculations.
DFT calculations with several Skyrme parameterizations predict a large jump in the charge radius around the neutron $N=104$ mid shell, 
with an odd-even staggering pattern related to the coexistence of nearly-degenerate oblate and prolate minima. 
This near-degeneracy is highly sensitive to many aspects of the effective interaction, 
a fact that renders perfect agreement with experiment out of reach for current functionals.
Despite this inherent difficulty, the SLy5s1 and a modified UNEDF1$^{SO}$ parameterization
predict a qualitatively correct staggering that is off by two neutron numbers. 
MCSM calculations of states with the experimental spins and parities show good agreement
for both electromagnetic moments and the observed charge radii. A clear mechanism for 
the origin of shape staggering within this context is identified: a substantial
change in occupancy of the proton $\pi h_{9/2}$ and neutron $\nu i_{13/2}$ orbitals.


\end{abstract}
\pacs{Valid PACS appear here}
\maketitle
\section{\label{sec:Intro}Introduction}
More than four decades ago, an unexpected large difference in the mean-square charge radius between $^{187}$Hg and $^{185}$Hg was observed by measuring 
the isotope shift in a Radiation Detection of Optical Pumping (RADOP) experiment performed at ISOLDE \cite{1972_Bonn_PLB,Bonn1976}. 
Similar to $^{185}$Hg, the $^{181,183}$Hg isotopes were found to exhibit a large isotope shift 
from their even-mass neighbours $^{182,184,186}$Hg \cite{Kuhl1977,Ulm1986}. 
Ever since these measurements, the observed pattern became known as `shape staggering'.
Studying the levels at low excitation energy in more detail, different shapes were identified in close vicinity to the ground state and the mercury isotopes are now one of the most illustrative examples of shape coexistence \cite{2011_Heyde_Wood_RevModPhys}. 
The experimental findings sparked extensive interest in studying this region of the nuclear chart 
from both experimental and theoretical points of view \cite{2011_Heyde_Wood_RevModPhys}. 
The large radius staggering was interpreted as transitions between weakly-deformed, oblate ground states and strongly-deformed, prolate ground states \cite{Frauendorf1975}.

The isotopic chain of mercury has since been studied with a multitude of complementary techniques:  
Coulomb excitation \cite{Bree2014,2016_Kasia_Liam_JPhysG,KasianickInProgress2017},
in-beam gamma-ray spectroscopy with recoil-decay tagging \cite{2001_Julin_intruderstates_RDT,2002_Kondev_PLB_179Hg_RDT,2003_melerangi_PRC_177Hg_RDT,ODonnel2012},
mass measurements \cite{2001_schwarz_ISOLTRAP_hg_masses} and $\alpha$/$\beta$-decay spectroscopy \cite{Cole1984,2002_jenkins_RDT_179Hg,Elseviers2011,Sauvage2013,RapisardaInProgress2017,Andreyev2009,Andreyev2010}. 
However, isotope shift and hyperfine-structure measurements had only been extended down to $^{181}$Hg \cite{Bonn1976,Ulm1986}.
While the ground-state deformation 
has been indirectly inferred for neutron-deficient mercury isotopes from in-beam recoil-decay tagging measurements, 
hinting towards less-deformed shapes for A$<$180 \cite{2001_Julin_intruderstates_RDT,2002_Kondev_PLB_179Hg_RDT,2003_melerangi_PRC_177Hg_RDT,ODonnel2012}, this had not been confirmed by a direct ground-state isotope-shift measurement.
The missing mean-square charge-radii data for the lighter mercury isotopes left the key question of where the shape staggering ends.

In order to address this key question, a measurement campaign was undertaken at the radioactive ion-beam facility ISOLDE \cite{Richard2017} performing in-source laser resonance-ionization spectroscopy 
of 15 mercury isotopes, ranging from the neutron-deficient to the neutron-rich side ($^{177-185,198,202,203,206-208}$Hg) 
with the goal of measuring their isotope/isomer shifts (IS) and hyperfine structures (HFS). 
The large isotopic span was made possible by 
using the Resonance Ionization Laser Ion Source (RILIS) \cite{2017Fedoseev} in a novel target-ion source combination \cite{2016_TDG_NIM_vadlis}, 
together with three different ion-counting techniques tailored to the isotope under investigation \cite{2013Marsh}: 
$\alpha$-decay spectroscopy for short-lived isotopes with small production rates (down to 0.1 ion/s) 
using a `Windmill'-type implantation station (WM) \cite{Andrei2010, Seliverstof2014}, 
Multi-Reflection Time-of-Flight mass spectrometer/separator (MR-ToF MS) \cite{2013Kreim} for high-resolution, single-ion counting and 
Faraday Cup (FC) ion-current measurements for high-intensity \mbox{($>$1 pA)} mercury beams. 
This paper is an in-depth follow-up article of \cite{Bruce_nature} on the neutron-deficient isotopes $^{177-185}$Hg.
A dedicated paper will provide a detailed discussion of the neutron-rich isotopes \cite{2017_paperTOM} also measured in the same experimental campaign.


\begin{figure}[hb]
\centering
\medskip
\includegraphics[width=1.0\columnwidth]{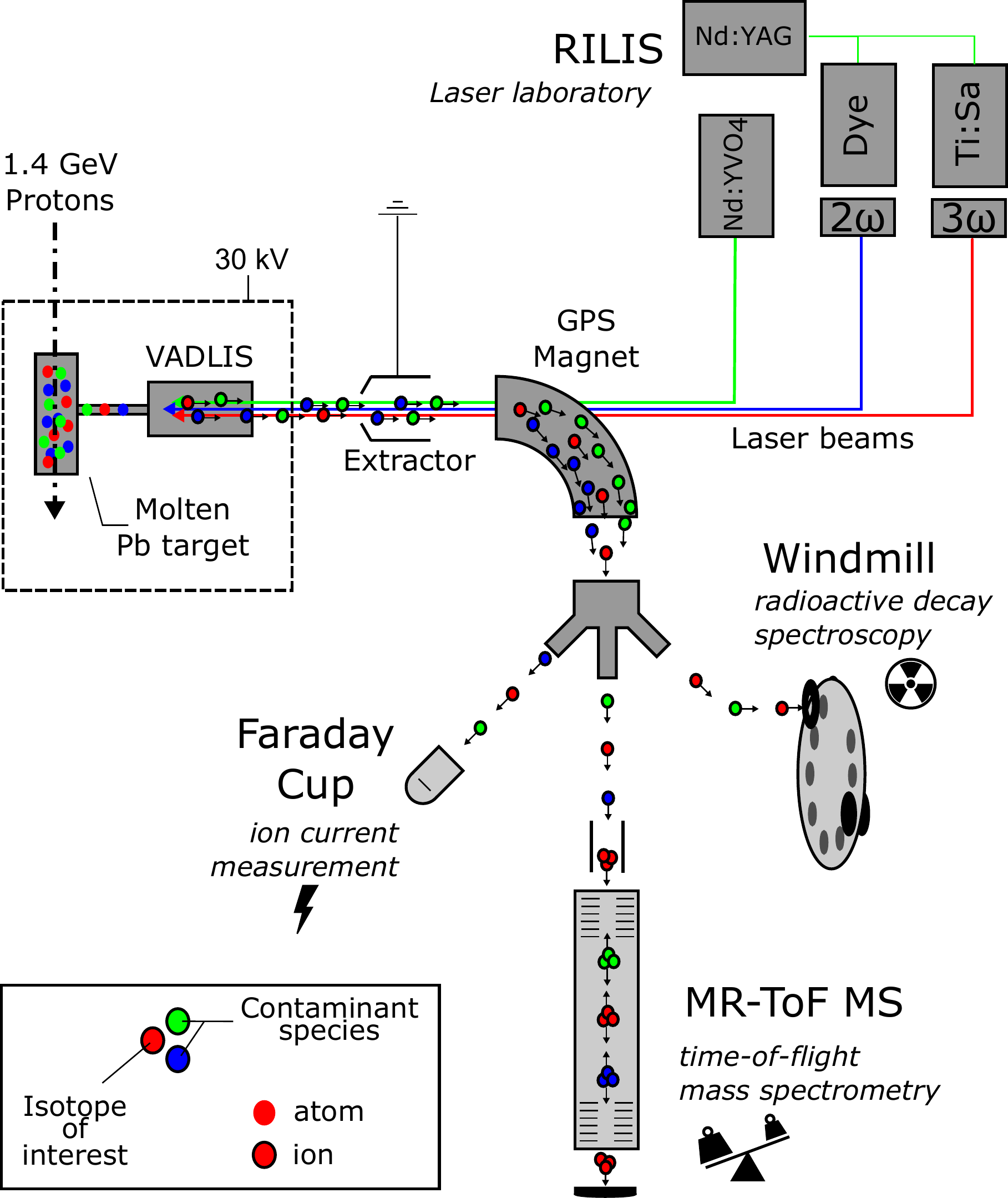}
\caption{Schematic of the experimental setup in the ISOLDE facility at CERN. See text for details.}
\label{fig:setup} 
\end{figure}

\begin{figure}[ht]
\centering
\medskip
\includegraphics[width=1.0\columnwidth]{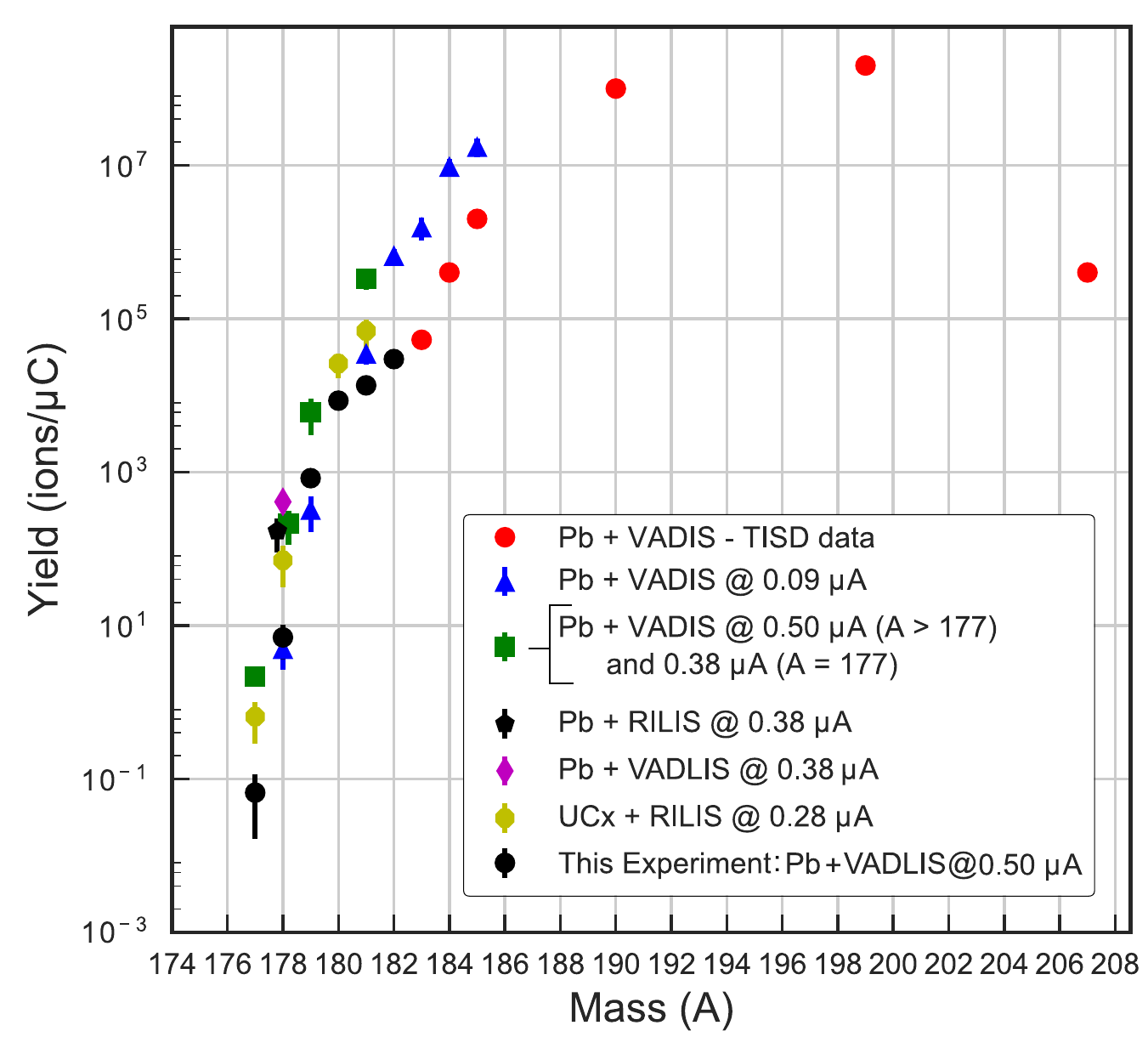}
\caption{Mercury production yield data for different target-ion source configurations: Versatile Arc Discharge Laser Ion Source (VADLIS) or Resonance Ionization Laser Ion Source (RILIS) with lead or uranium-carbide target material for different proton-beam currents.}
\label{fig:Hg_yields}
\end{figure}

\section{\label{sec:Experiment}Experimental Technique}
\subsection{\label{sec:Production}Mercury ion beam production}
Mercury isotopes were produced at the CERN-ISOLDE facility \cite{Richard2017} via spallation reactions induced by a 1.4-GeV proton beam from the PS-Booster synchrotron impinging upon a molten-lead target.
The neutral reaction products effused from the heated target via the transfer line ($\approx$700 $^{\circ}$C target and $\approx$400 $^{\circ}$C transfer line heating) into the VADLIS cavity \cite{2016_TDG_NIM_vadlis}, which was operated in RILIS mode \cite{2016_TDG_NIM_vadlis} (Fig. \ref{fig:setup}). 
In this mode, lasers are used to resonantly ionize the isotopes of interest.
The photo-ions were extracted and accelerated by a \mbox{30 kV} potential difference, mass separated by ISOLDE's General Purpose Separator (GPS) dipole bending magnet before being sent to one of three measurement devices (FC/WM/MR-ToF MS; see Fig. \ref{fig:setup}).
The choice of a molten-lead target was based on results obtained from a preparatory experiment in similar conditions at ISOLDE, where the mercury production of a molten-lead was compared with a UC$_x$ target (Fig. \ref{fig:Hg_yields}). 
While the production rates of the lightest mercury isotopes were of a similar order of magnitude for both cases, the use of a molten-lead target significantly reduces the isobaric contamination of surface-ionised contaminants. 
This was especially important for the heavy mass region discussed in \cite{2017_paperTOM}.
Furthermore, the production rate of the heavy mercury isotopes was significantly higher for the molten-lead target.

\begin{figure}[ht]
\centering
\medskip
\includegraphics[width=1.0\columnwidth]{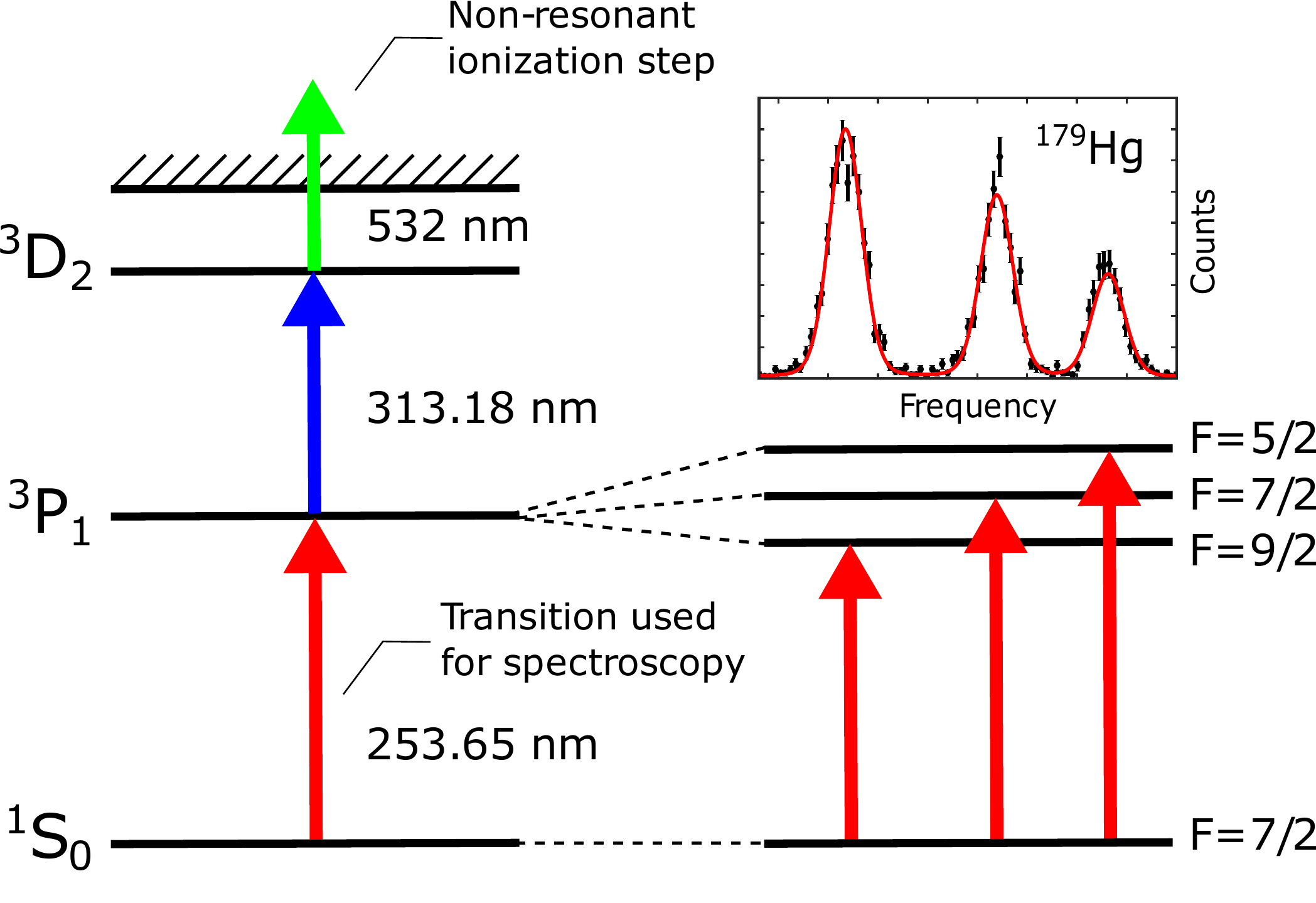}
\caption{Ionization scheme for mercury developed in \cite{DayGoodacre_Hg_efficiency}, using the same transition for spectroscopy as in \cite{1972_Bonn_PLB} and \cite{1975_Bonn_ZPhysA}. The right-hand side shows the splitting of the $^{3}\!P_1 $ level states with different total angular momentum $F=I+J$ for an $I=7/2$ nuclear spin with a corresponding exemplary HFS spectrum of $^{179}$Hg. Energies and level splitting are not displayed to scale.}
\label{fig:laserschema}
\end{figure}

Resonance ionization of the mercury isotopes was accomplished using a 3-step ionization scheme (Fig. \ref{fig:laserschema}) with a measured ionization efficiency of 6$\%$  \cite{DayGoodacre_Hg_efficiency}. 
Laser spectroscopy was performed on the \mbox{253.65-nm} $ 6s^{2}~^{1} S_0 \rightarrow 6s6p ^{3} P_1 $ transition.
Well-resolved HFS spectra were obtained by scanning the frequency-tripled wavelength of the Ti:Sapphire laser with a bandwidth of $\approx$1.5-GHz FWHM after tripling (labelled $3\omega$ in Fig. \ref{fig:setup}).
At the second step (313.18 nm) the frequency-doubled output of the dye laser (Credo Dye model by Sirah Lasertechnik GmbH \cite{Credo}) was used.
The third step was a non-resonant 532-nm transition driven by a Nd:YVO$_4$ laser.

\subsection{\label{sec:Counting}Isotope identification and counting}
\subsubsection{\label{sec:FaradayCup}Ion-current measurement with a Faraday cup}
A Faraday cup installed downstream the GPS magnet was used to measure the extracted ion current of sufficiently intense 
($>$1 pA) beams of the longer-lived mercury isotopes. 
In this experiment, the reference isotope for IS measurements, stable $^{198}$Hg, as well as radioactive $^{202,203}$Hg were probed using this technique. 
In the case of $^{198}$Hg, contamination of the neighboring $^{197}$Hg ground and isomeric states was present in the HFS.
To prove consistency of the IS measurements for different techniques, all isotopes measured using the FC were also measured with the MR-ToF MS technique, 
with which it was possible to suppress the unwanted isotopic and isobaric contaminant species.

\subsubsection{\label{sec:Windmill}IS and HFS measurements with the Windmill setup}

The Windmill detection setup \cite{Andrei2010,Seliverstof2014}, consists of a vacuum chamber holding a rotatable wheel that houses 10 thin carbon foils up to 12-mm diameter (20 $\mu$g/cm$^2$ thickness) \cite{2002_Lommel_NIMA}. 
Surrounding these foils are two pairs of silicon detectors. 
The first pair is positioned around the carbon foil in which the beam is implanted. 
This pair consists of an annular (Ortec, TC-025-450-300-S, 6 mm hole, 450mm$^2$ active area) and a full surface barrier detector (Ortec, TB-020-300-500, 300mm$^2$ active area).
The beam is implanted through the central hole of the annular detector.
After implantation, the wheel rotates and a fresh foil is placed in view of the beam. 
The foil that was irradiated is moved towards the so-called `decay position' between a second pair of silicon detectors.
The two silicon detectors (Canberra, PD 300-15-300 RM, 300mm$^2$ active area) were used to study longer-lived radioactive isotopes that were
implanted directly or the daughter products of previously implanted nuclei. 
The total $\alpha$-particle detection efficiency at the implantation site was 34$\%$, with a detector resolution of 35-keV FWHM.
Two germanium detectors were additionally positioned outside the chamber of the WM setup, 
in order to measure $\gamma$- and X-ray radiation emitted from the implantation site.\\

\clearpage

\begin{figure}[h]
\centering
\includegraphics[width=1.0\columnwidth]{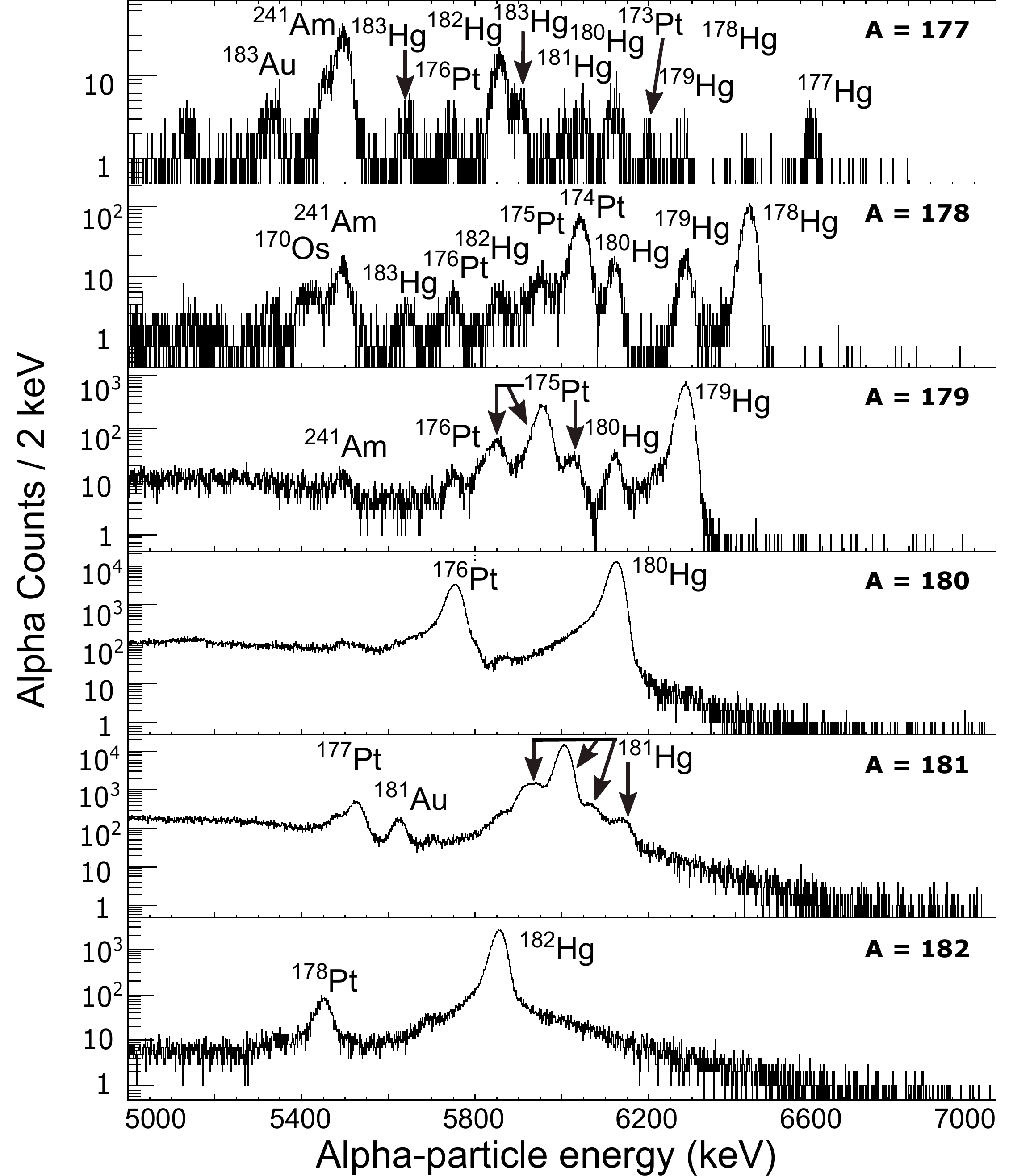}
\caption{$\alpha$-particle energy spectra obtained at implantation site when the GPS was set to masses in the range of $A=177-182$.
Because of the limited mass resolution and difference in production, beam contaminants of heavier mercury isotopes are observed in the lighter-mass spectra.}
\label{fig:Alpha_spectra_and_tof_spectra_A}
\end{figure}

\begin{figure}[h]
\centering
\medskip
\includegraphics[width=1.0\columnwidth]{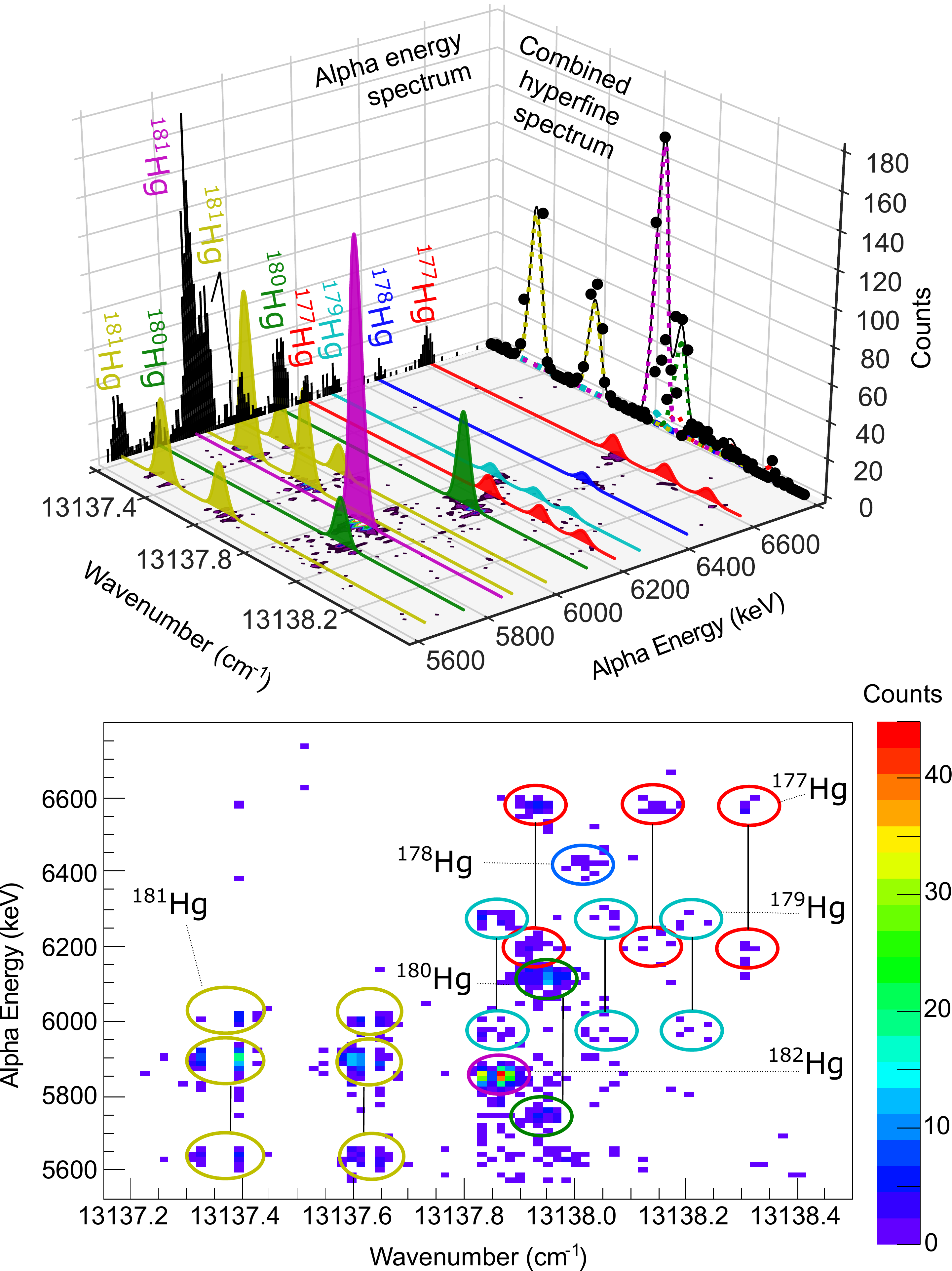}
\caption{ Top: 3D-spectrum showing counts as function of $\alpha$-decay energy and laser frequency before tripling (in wavenumber).
The spectrum is obtained from a single laser scan measured at the Windmill implantation site when the GPS mass separator was set to $A = 177$.
Because of the limited mass resolution and approximately four orders of magnitude larger production rate of $^{182}$Hg in comparison to $^{177}$Hg, 
beam contaminants are visible up to $A = 182$. 
The HFS for each contaminant is shown by the colored lines along the Z-axis, where different colors represent different mercury isotopes.
When projecting on the $\alpha$ energy or laser-frequency axis, the total $\alpha$-decay energy spectrum and combined HFS for all isotopes appear respectively.
Bottom: The projection of the top plot on the $\alpha$-decay energy and laser-frequency plane shows regions of counts (indicated by circles) related to the isotopes decay and HFS.
Lines drawn between circles indicate the same HFS peaks visible in daughter products of the original decaying isotopes.}
\label{fig:Alpha_Freq_3D_spectra}
\end{figure}

The $\alpha$-particle energy spectra obtained at the implantation point for different mass-separator settings are shown in Fig. \ref{fig:Alpha_spectra_and_tof_spectra_A}.
By using the $\alpha$-particle energies to identify short-lived isotopes and associating the count rates with the wavenumber of the laser targeting the spectroscopic transition, it is possible to produce nearly-background-free HFS when the $\alpha$ particles coming from the decay of the beam contaminants have sufficiently differing energies. 
This is for instance shown in the $\alpha$-energy spectrum collected with $A = 177$ mass-separator setting where contaminants from $^{178-185}$Hg were present in the beam (Fig. \ref{fig:Alpha_Freq_3D_spectra}). 
Here, clean HFS of 6 mercury isotopes can be observed in a single scan simply by gating on different $\alpha$ peaks. 
The efficiency and selectivity of this method allows IS and HFS measurements of isotopes with very small production rates such as $^{177}$Hg which was delivered at a rate of $\approx$0.1 ions s$^{-1}$. 

\vspace{30pt}

\clearpage
\subsubsection{\label{sec:MRTOF}IS and HFS measurements with \mbox{Multi-Reflection-Time-Of-Flight technique}} 
For the measurement of $^{183-185,198,202,203,206-208}$Hg, ISOLTRAP's MR-ToF MS \cite{2013Kreim} was used for mass separation and ion detection. 
This device, extensively discussed in \cite{2013_Wolf_IJMS_MRTOF}, consists of two 160 mm-long, 6-fold electrostatic mirrors surrounded by shielding electrodes. 

\begin{figure}[h]
\centering
\includegraphics[width=1.0\columnwidth]{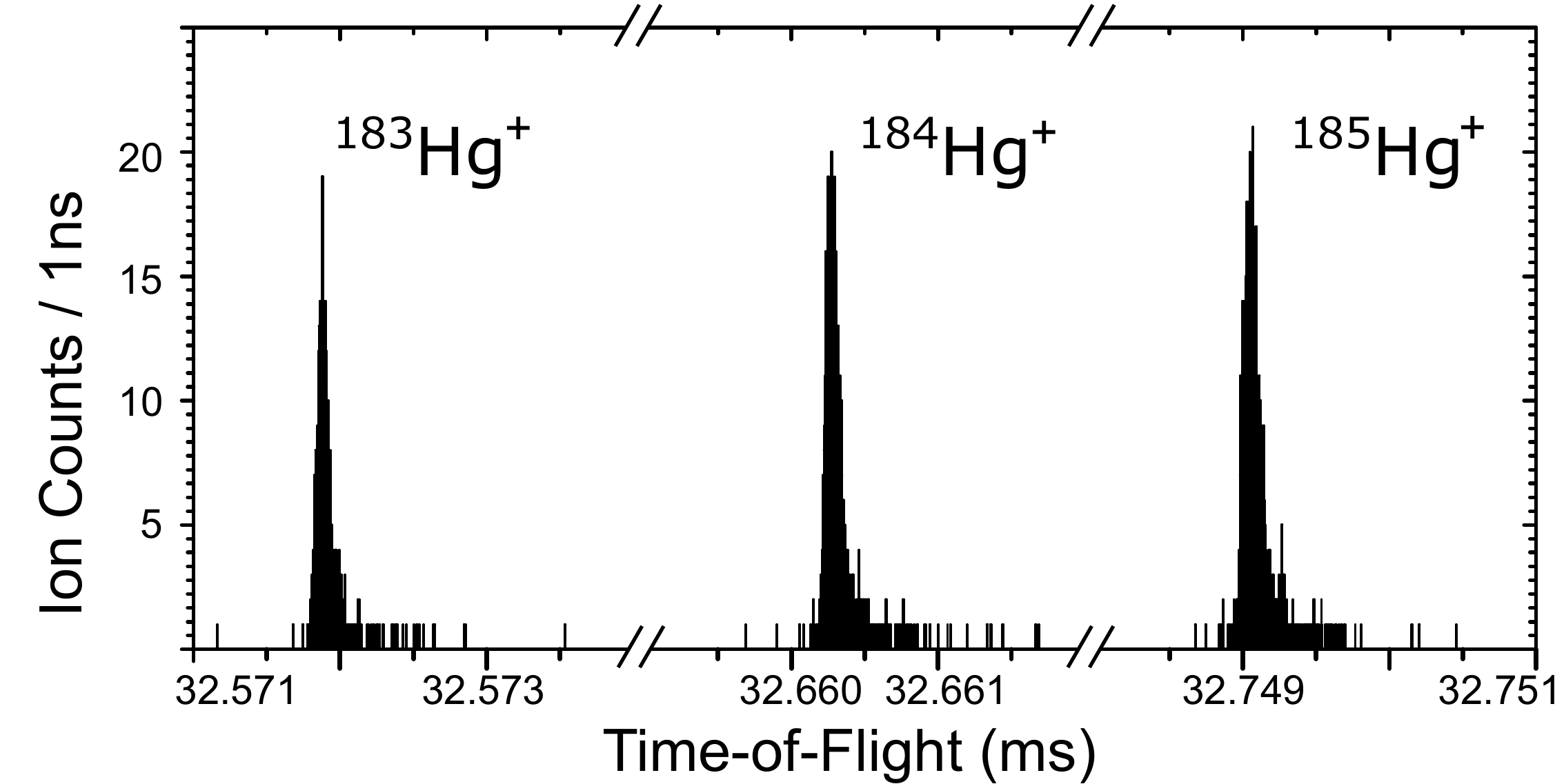}
\caption{Time-of-Flight spectra obtained for $^{183-185}$Hg, when the GPS mass separator was set to $A = 183, A = 184, A = 185$ respectively, after 1000 revolutions in the MR-ToF MS at which a mass resolving power of about 1.2$\times 10^5$ was reached. At this resolving power, all isotopic and isobaric contaminants are cleared from the spectra.}
\label{fig:Alpha_spectra_and_tof_spectra_B}
\end{figure}

First, the ion beam from ISOLDE is injected into a radio-frequency quadrupole cooler-buncher (RFQCB) \cite{HERFURTH2001254}.
Ion bunches are stopped and thermalized in this helium-filled RFQCB before they are injected into the MR-ToF MS with at typical energy spread of 60 eV and
bunch width of 60 ns \cite{2013Kreim,Cubiss2018}.
Here, the ion bunches are trapped by reducing their kinetic energy and switching of the in-trap lift voltage. 
In the trapping cavity, they undergo multiple round-trips between the electrostatic mirrors, 
where the time-of-flight is dependent on the mass and the charge state.
This causes a separation in time for different isobaric species in each bunch.
The ion bunches are then ejected from the MR-ToF MS by switching the in-trap lift voltage \cite{2013Kreim}.
The arrival time of the ejected, mass-separated ion bunches was measured by employing a MagnetTOF$^{\text{TM}}$ 
secondary electron multiplier ion detector (DM291, ETP, Ermington, Australia).
This MR-TOF mass separator is able to reach resolving powers of $R=\Delta m/m =10^5$ within a few ten milliseconds \cite{2013Kreim}.
The procedure for employing the MR-ToF MS in a laser-spectroscopy experiment was previously discussed in \cite{Cubiss2018}.
Examples of different time-of-flight spectra are shown in Fig. \ref{fig:Alpha_spectra_and_tof_spectra_B}.

\subsection{\label{sec:HFS construction}Recording of HFS}
\begin{figure}[h]
\centering
\medskip
\includegraphics[width=1.0\columnwidth]{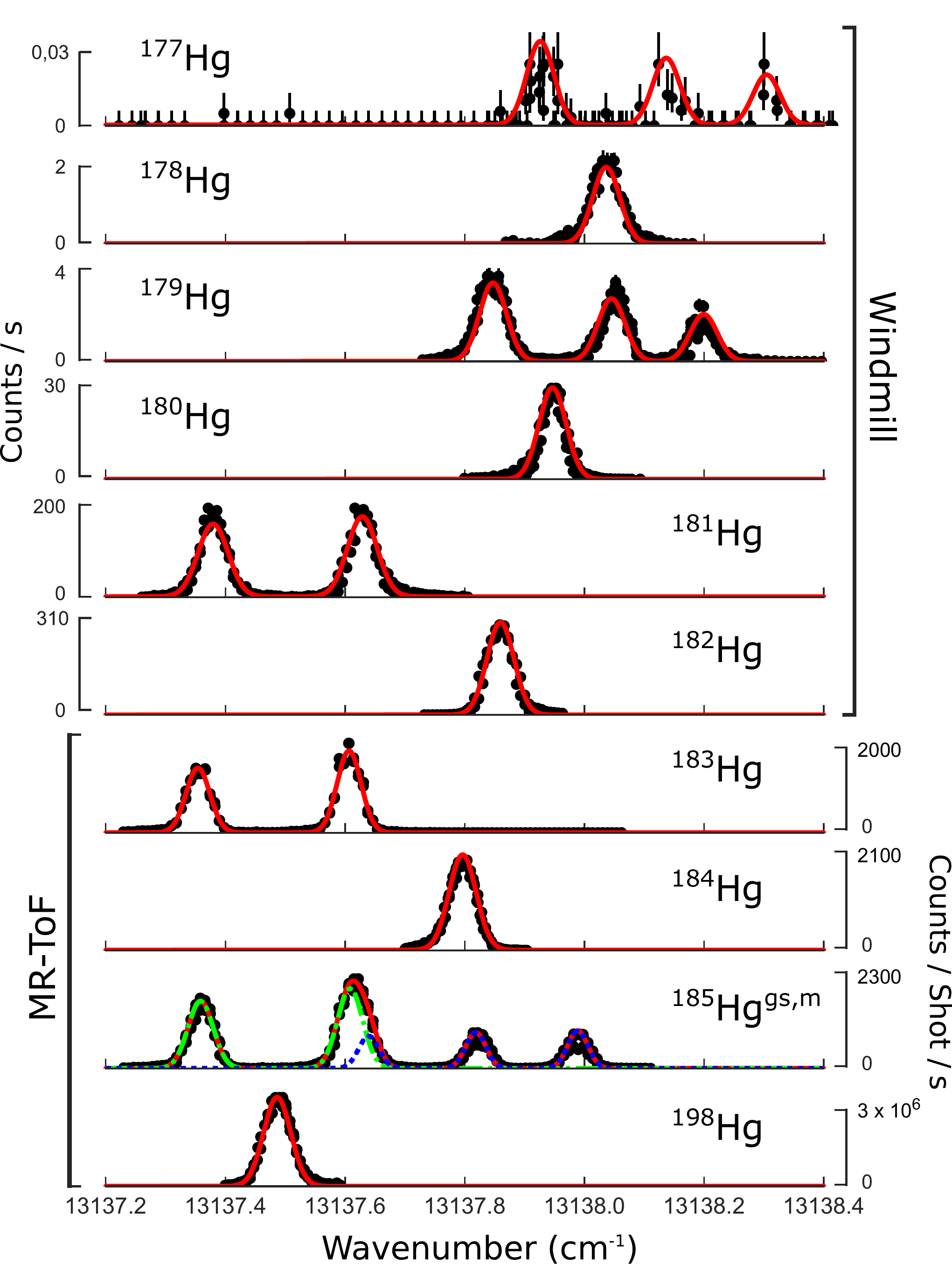}
\caption{Examples of the HFS scans taken for the isotopes discussed in this paper. 
A fit of the data is shown in red. For $^{185}$Hg, the green and blue dotted lines correspond to the ground and isomeric state parts of the total fit, shown in red, respectively.}
\label{fig:Data_summary}
\end{figure}

During the experiment, the wavelength of the laser targeting the spectroscopic transition was scanned in a step-wise manner and a parameter proportional to the number of detected photo-ions was recorded as a function of the scanned laser frequency.
The data taking was synchronized with ISOLDE's super-cycle structure of proton pulses provided by the Proton Synchrotron Booster. 
Depending on the production rate of the isotope under investigation, between 1 and 5 full super-cycles of measurement were taken for each laser frequency step. 
Data obtained with the WM setup were recorded with an event-by-event data structure. The data were analyzed off-line with the ROOT software package \cite{ROOT}, 
where the deadtime-corrected integral of energy-gated alpha counts for each laser frequency resulted in the hyperfine spectra. 
A similar analysis was performed for the MR-ToF MS data where isotope counting was not based on energy of an emitted particle but on ion arrival time at the ion detector. 
For measurements with the Faraday cup, the integrated ion current for each laser step was combined with the recorded laser frequency to create the HFS.
An overview of the measured HFS is given in Fig. \ref{fig:Data_summary}.

\clearpage
\onecolumngrid

\begin{table*}[h]
\centering
\caption{Isotope shifts ($\delta \nu^{198,A}$) and hyperfine splitting parameters ($a$ and $b$) for the $6s6p ^{3}\!P_1$ level in mercury atoms and \\
deduced mean-square charge radii ($\delta \langle r^2\ \rangle ^{198,A}$) and electromagnetic moments ($\mu$ and $Q$) in $^{177-185}$Hg istopes.\\
Results of both optional ground-state spin assigments $I=7/2$ and $I=9/2$ for $^{177,179}$Hg are shown (see Sec. \ref{sec:moments}).\\
The literature data for $\delta \langle r^2\ \rangle ^{198,A}$ in this table are recalculated from the experimental IS \cite{Ulm1986}.}
\label{table:hyperfine}
\begin{ruledtabular}
\begin{tabularx}{\textwidth}{rrccclccc}
\multicolumn{1}{c}{Isotope} & \multicolumn{1}{c}{$I^{\pi}$ } & \multicolumn{1}{c}{ $\delta \nu^{198,A}$ } & \multicolumn{1}{c}{$a$} 
&\multicolumn{1}{c}{$b$} & \multicolumn{1}{c}{$\delta \langle r^2\ \rangle ^{198,A}$ $^{a}$ } & \multicolumn{1}{c}{$\mu$} & \multicolumn{1}{c}{$Q_s$} & \multicolumn{1}{c}{Ref}\\

\multicolumn{1}{c}{} & \multicolumn{1}{c}{} & \multicolumn{1}{c}{(MHz)}	&\multicolumn{1}{c}{(MHz)} & \multicolumn{1}{c}{(MHz)} & \multicolumn{1}{c}{(fm$^2$)}  
 & \multicolumn{1}{c}{($\mu_N$)} & \multicolumn{1}{c}{($b$)} & \multicolumn{1}{c}{}\\

\hline
\\
$^{177}\mathrm{Hg}$  & $(7/2^-)$ & 54580(390)   & -4320(180) & -410(600)  & -1.067(8)\{78\}  & -1.025(48)$^{b}$ & 0.57(83) & this work\\
					 & $(9/2^-)$ & 55170(390)	& -3460(180) & -875(600)  & -1.083(8)\{78\}  & -1.035(60)$^{b}$ & 1.21(91) & this work\\
$^{178}\mathrm{Hg}$  & $0^+$     & 49500(290)   & -          & -          & -0.968(6)\{71\}  & - 		 & - 	     & this work\\
$^{179}\mathrm{Hg}$  & $(7/2^-)$ & 46310(240)   & -3990(80)  & -550(200)  & -0.905(5)\{70\}  & -0.948(24)$^{b}$ & 0.76(28) & this work\\
  					 & $(9/2^-)$ & 46820(230)   & -3150(70)  & -1050(210) & -0.915(5)\{70\}  & -0.947(27)$^{b}$  & 1.45(31) & this work\\
$^{180}\mathrm{Hg}$  & $0^+$     & 41330(240)   & -          & -          & -0.808(5)\{60\}  & - 		 & - 	      & this work\\
$^{181}\mathrm{Hg}$  & $1/2^-$   &  5390(280)   & 15030(120) & - 		  & -0.111(6)\{11\}  & 0.515(4)  & - 	      & this work\\
					 & 			 &  5560(200)   & 14960(250) & -  		  & -0.114(4)\{10\}  & 0.5071(7) & -	      & \cite{Bonn1976,Ulm1986}\\
$^{182}\mathrm{Hg}$  & $0^+$     & 33350(260)   & -          & - 		  & -0.653(5)\{48\}  & -         & - 	      & this work\\
$^{183}\mathrm{Hg}$  & $1/2^-$   &  3100(260)   & 15190(160) & - 		  & -0.065(5)\{7\}   & 0.521(6)  & - 	      & this work\\
					 &			 &  3310(100)   & 15380(130)  & - 		  & -0.069(2)\{6\}   & 0.524(5)  & - 	      & \cite{Bonn1976,Ulm1986}\\
$^{184}\mathrm{Hg}$  & $0^+$     & 27680(270)   & -          & - 		  & -0.542(6)\{40\}  & - 		 & - 	      & this work\\
					 & 	 		 & 27720(90)    & -          & - 		  & -0.544(2)\{42\}  & - 		 & - 	      & \cite{Ulm1986}\\
$^{185}\mathrm{Hg}$  & $1/2^-$   & 3350(300)    & 14930(340) & -  		  & -0.069(6)\{7\}   & 0.51(1)   & - 	      & this work\\
					 & 			 & 3710(30)     & 14960(70)  & -  		  & -0.0764(6)\{63\} & 0.509(4)  & - 	      & \cite{Ulm1986}\\
$^{185}\mathrm{Hg}^{m}$ & $13/2^+$  & 27780(190)   & -2286(25)  &  110(300)  & -0.543(4)\{40\} & -1.01(1)   & -0.15(41)& this work\\
					 & 			 & 27770(110)   & -2305(19)  & -140(230)  & -0.543(2)\{42\}  & -1.017(9) &  0.20(33)& \cite{Ulm1986}\\
\\
\hline
\multicolumn{9}{l}{$^{a}$ Statistical errors are given in parenthesis. Systematic errors stemming from the indeterminacy of the }\\
\multicolumn{9}{l}{\ \ $F$ factor (7\%) \cite{Ulm1986} and $M_{\text{SMS}}$ are shown in curly brackets (see Eqs. \ref{eq:dr}-\ref{eq:dr2})}\\
\multicolumn{9}{l}{$^{b}$ Corrected in accordance with hyperfine anomaly estimation (see Sec. \ref{sec:momnents})}

\end{tabularx}
\end{ruledtabular}
\end{table*}

\twocolumngrid

\section{\label{sec:Results} Results}
\subsection{\label{sec:Fitting}Extraction of IS and hyperfine splitting parameters}

Information on the difference in mean-square charge radius \mbox{$\delta\langle r^2\ \rangle ^{A,A'}$} between two nuclei with mass $A$ and $A'$ of the same isotopic chain 
is extracted from the difference in the positions of the centers of gravity of their respective HFS, $\nu_0$ i.e. their isotope shift of a certain transition.
The nuclear electromagnetic moments (magnetic dipole, electric quadrupole) dictate the relative position of the an atomic-state hyperfine-splitting component with respect 
to $\nu_0$ via the relation
\begin{equation}
\Delta \nu^F = 0.5 a K + b \frac{0.75 K (K+1)- I(I+1) J(J+1)}{2IJ(2I-1)(2J-1)},
\end{equation}
where the dipole and quadrupole hyperfine splitting parameters are given as $a$ and $b$, $\Delta \nu^F$ represents the energy difference of the hyperfine component with total angular momentum $\mathbf{F=I+J}$, with respect to $\nu_0$ \cite{Otten1989} and $K = F(F+1)-I(I+1)-J(J+1)$.
Fitting of the spectra was performed with the open-source Python package SATLAS \cite{SATLAS} and cross-checked with a similar fitting routine in ROOT \cite{ROOT} and the fitting procedure that was used in our previous HFS studies as for instance in \cite{Seliverstof2014}.

To monitor the stability of the whole system, reference scans of $^{198}$Hg were performed regularly.
The spectra were fitted separately and the weigthed mean of the fit results is taken as a final value. 
Results of the fits are shown in table \ref{table:hyperfine}.
The experimental errors on the IS include both the fit errors and the spread of individual scan results. 
For $^{177}$Hg, where only a single full spectrum was obtained, the typical dispersion in the extracted HFS centroid position for the other isotopes was added as an additional uncertainty.
As the nuclear spin of $^{177,179}$Hg could not be determined directly by counting the hyperfine components from the present measurements due to the low angular momentum of the electronic state ($J=1$) of the upper level of the studied transition (see Fig. \ref{fig:laserschema}), we report the IS and hyperfine splitting constant values assuming both possible options for the ground-state nuclear spin of $^{177,179}$Hg (7/2 and 9/2, see section \ref{sec:moments}).
Within the experimental uncertainties, good agreement on the IS, $a$ and $b$ parameters compared to the previous measurements was obtained.


\subsection{\label{sec:radii} Changes in mean-square charge radii}

The isotope shift $\delta \nu_i^{A,A'}$ between two isotopes of the same isotopic chain with mass $A$ and $A'$ for transition $i$, results from the Mass and Field shifts noted $\delta \nu_{M,i}^{A,A'}$ and $\delta \nu_{F,i}^{A,A'}$ respectively:
\begin{equation}\label{eq:dr}
\delta \nu_i^{A,A'} = \nu_i^{A'} - \nu_i^{A} = \delta \nu_{M,i}^{A,A'} + \delta \nu_{F,i}^{A,A'}.
\end{equation}
The mass shift can be described as the sum of the so-called normal (NMS) and specific (SMS) mass shifts:
\begin{equation}\label{eq:dr1}
\delta \nu_{M,i}^{A,A'} = M \frac{A'-A}{AA'} = (M_{\text{NMS}} +  M_{\text{SMS}})\frac{A'-A}{AA'},
\end{equation}
where the NMS is related to the ratio of the electron and proton masses $m_e$ and $m_p$ and to the transition frequency $\nu_{i}$ as 
$M_{\text{NMS}} = \frac{m_e}{m_p} \nu_{i}$.
The field shift is proportional to an electronic $F$ factor and nuclear parameter $\lambda^{A,A'}$, related to the change in nuclear mean-square charge radius between the two isotopes according to:
\begin{equation}\label{eq:dr2}
\delta \nu_{F,i}^{A,A'} = F_{\lambda,i} \ \lambda^{A,A'} = K(Z) F_{\lambda,i}  \delta \langle r^2\ \rangle ^{A,A'}.
\end{equation}
In this equation, $\lambda^{A,A'}$ takes into account the influence of the higher-order radial moments. 
It was shown \cite{1985_TorbohmFricke} that the difference between $\lambda$ and $\delta \langle r^2\ \rangle $ is small (less than 10\% for heavy atoms) and can be accounted for by the single correction factor $K(Z)$.
The $F$ and $M$ factors used, as well as the higher radial moments correction $K(Z)$ were taken from \cite{2004_FrickeHeilig}, resulting from a combined analysis of data from optical spectroscopy, muonic atoms and elastic electron scattering.
The used values are $F_{254\ \text{nm}} = -53(4)  \text{ GHz/fm}^2$, $M^{\text{SMS}}_{254 \ \text{nm}} = 0 \pm 0.5 M^{\text{NMS}}$ and \mbox{$K(Z)$ = 0.927}.\\

\begin{figure}[ht]
\centering
\medskip
\includegraphics[width=1.0\columnwidth]{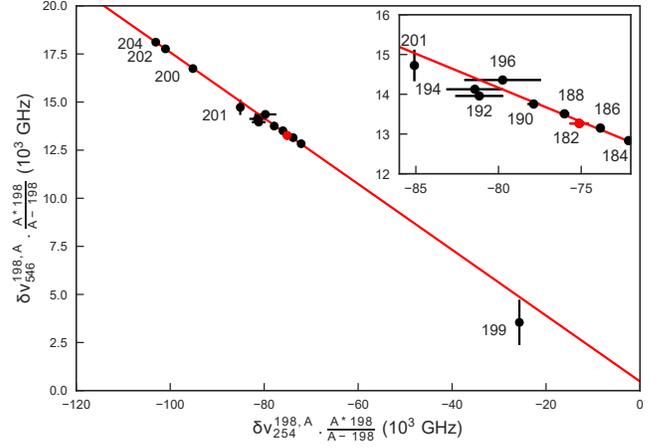}
\caption{King plot of the modified isotope shifts in the 546-nm line versus those in the 254-nm line with $^{198}$Hg as reference isotope. 
Inset: zoom around the area of the plot where most of the points are clustered. The new data point for $^{182}$Hg fits well to the King-plot line from \cite{Ulm1986}.}
\label{fig:King_Plot}
\end{figure}

A King plot combines IS for two different atomic transitions to extract information on the electronic $F$ and $M$ factors. 
The linear relation that exists between the modified isotope shifts ($\delta\nu_{i}^{A', A}.\frac{A A'}{A'-A}$) following from equations (\ref{eq:dr})-(\ref{eq:dr2}), has a slope $\kappa$ that equals to the $F$-factor ratio ($\kappa = F_i/F_{i'}$) of the two transitions $i$ and $i'$.
Information on the $M$ factors can be derived from the intercept with the y-axis, $s$, via $s = M_{i'} - \kappa M_{i}$.
In \cite{Ulm1986}, this procedure was used to determine $F$ and $M$ factors for the 546-nm transition from the fixed factors for the 254-nm transition.
The IS of $^{182}$Hg was measured in \cite{Ulm1986}, only for the 546-nm transition. To check the consistency of our data on IS for $^{182}$Hg in the 254-nm transition, we include the corresponding point into the King plot from \cite{Ulm1986} (see Fig. 8).
As can be seen in Fig. \ref{fig:King_Plot} and its inset, the data from this work matches the previously observed trend. \\

The isotope shifts obtained from fitting and the resulting calculated differences in mean-square charge radius $\delta \langle r^2\ \rangle^{A,A'}$ 
are shown in table \ref{table:hyperfine} and plotted in Fig. \ref{fig:Charge_radii_comp}.
From \mbox{$\delta \langle r^2\ \rangle ^{A,A'}$}, the mean-squared deformation parameter $\beta_2 =<\beta^2>^{1/2}$ can be inferred from the relation \cite{Otten1989}
\begin{equation}\label{eq:r2}
\langle r^2\ \rangle  = \langle r^2\ \rangle _{\text{DM}}^{\text{spher}} \left(1 + \frac{5}{4\pi}\langle \beta_2^2\ \rangle \right),
\end{equation}
where $\langle r^2\ \rangle _{\text{DM}}^{\text{spher}}$ represents the droplet model prediction for a spherical nucleus.
The droplet-model calculations have been carried out using the second parametrization of Berdichevsky and Tondeur \cite{Berdichevsky1985}.

\subsection{\label{sec:momnents}Electromagnetic moments}

The magnetic moments, $\mu $, of the discussed mercury isotopes were calculated using the relation
\begin{equation}
\mu_A = \mu_{A_0}.\frac{I_A}{I_{A_0}}.\frac{a_A}{a_{A_0}}.(1+ ^{A_0}\!\Delta^{A})
\end{equation}
where we use $^{199}$Hg$^m$ as a reference ($A_0 = 199$, $\mu_{A_0} = -1.0147(8)$ $\mu_N$ \cite{Reimann1973}, $a_{A_0} = -2298.3(2)$ MHz \cite{Stroke}). 
The hyperfine anomaly, $^{A_0}\!\Delta^{A}$, is defined as 
\begin{equation}
^{{A_1}}\!\Delta^{{A_2}} = \frac{a_{A_1}}{g_{I,{A_1}}}.\frac{g_{I,{A_2}}}{a_{A_2}} - 1
\end{equation}
where $g_I$ is the nuclear $g$ factor and the indices $A_1$ and $A_2$ refer to two different isotopes with atomic mass numbers $A_1$ and $A_2$. The hyperfine anomaly arises from the differences in charge and magnetization distribution within the nucleus, through the 
``Breit-Rosenthal'' (BR) \cite{Breit1932} and ``Bohr-Weisskopf'' (BW) \cite{Bohr1950} effects, respectively. If the magnetic hyperfine constant for the point-like nucleus is denoted as $a_{\text{point}}$, then the observed magnetic hyperfine constant $a$ can be presented as follows:
\begin{equation}
a = a_{\text{point}}(1+\epsilon)(1+\delta)
\end{equation}
where $\epsilon$ and $\delta$ are responsible for the BW and BR effects, respectively. Then, the hyperfine anomaly acquires the simple expression:
\begin{equation}
^{A_1}\!\Delta^{A_2} = ^{A_1}\!\Delta^{A_2}_{\text{BW}}+^{A_1}\!\Delta^{A_2}_{\text{BR}} = (\epsilon_1 - \epsilon_2)+(\delta_1 - \delta_2).
\end{equation}
To determine the hyperfine anomaly one should have independent values for magnetic moments and $a$-constants for the pair of isotopes under study, measured with high accuracy. 
In the case of the mercury isotopes, such measurements were done earlier for ten long-lived isotopes and isomers ($^{193-201, 193m-199m}$Hg). Correspondingly, for these nuclei, the hyperfine anomaly is known with sufficient accuracy \cite{Persson2013}. 
Moskowitz and Lombardi \cite{Moskowitz1973} have shown that if one neglects the BR part of the anomaly in comparison with its BW component, then the experimental hyperfine anomalies in the $^{3}\!P_1$ atomic state of this series of mercury isotopes, with the odd neutron in the 
nuclear shell model neutron orbitals $p_{1/2}, p_{3/2}, f_{5/2}$ and $i_{13/2}$, are well reproduced assuming the simple relation known as the Moskowitz-Lombardi (ML) rule:
\begin{equation}
^{A_1}\!\Delta^{A_2}_{\text{BW}} = \pm \alpha \left( \frac{1}{\mu_1} - \frac{1}{\mu_2} \right), 
\end{equation}
with sign from $I= \ell \pm \frac{1}{2}$, where $\alpha= 1\times 10^{-2} \ \mu_N$ and $\ell$ is the orbital moment of the unpaired neutron.
Neglecting the BR part of the hyperfine anomaly was justified in \cite{Rosenberg1972} where $^{A_1}\!\Delta^{A_2}_{\text{BR}}$ was calculated with a diffuse nuclear charge distribution. 
In particular, according to \cite{Rosenberg1972}, $^{199}\!\Delta^{201}_{\text{BR}} = -1.8\times 10^{-4}$ whereas $^{199}\!\Delta^{201} = -1.5\times 10^{-3}$ \cite{Persson2013}. 
The ML-rule was further supported by calculations from a microscopic theory \cite{Fujita1975}. 
We applied the ML-rule to estimate the BW-correction for the magnetic moments of $^{177, 179}$Hg, taking into account the description of the experimental hyperfine anomaly 
by this rule for the variety of neutron single-particle states in mercury nuclei with the mass spanning a rather large range. 
For previously-measured isotopes and isomers the maximal deviation of the experimental $^{199}\!\Delta^{A}_{\text{BW}}$ from the ML-calculation is equal to $2.5\times10^{-3}$. 
We conservatively estimated the error of ML-prediction for the hyperfine anomaly in $^{177, 179}$Hg as $5\times10^{-3}$.
It was shown in \cite{Pendrill1995} that $^{A_1}\!\Delta^{A_2}_{\text{BR}}$ is proportional to \mbox{$\delta \langle r^2\ \rangle ^{A_1,A_2}$}. 
Thus, $^{199}\!\Delta^{A}_{\text{BR}}$ for $^{177, 179}$Hg can be estimated by scaling the calculated $^{199}\!\Delta^{201}_{\text{BR}}$ \cite{Rosenberg1972}.
The uncertainty of this correction was estimated to be 10$\%$.
For $^{177,179}$Hg, the BR correction are $^{199}\!\Delta^{177}_{\text{BR}}=-1.7\times10^{-3}$, $^{199}\!\Delta^{179}_{\text{BR}}=-1.5\times10^{-3}$. 
Assuming an $I=7/2$ assignment, 
$^{199}\!\Delta^{177}_{\text{BW}}=-1.2\times10^{-4}$ and  $^{199}\!\Delta^{179}_{\text{BW}}=7\times10^{-4}$. 
Under the $I=9/2$ assignment, $^{199}\!\Delta^{177}_{\text{BW}}=-1.9\times10^{-2}$ and $^{199}\!\Delta^{179}_{\text{BW}}=-2.0\times10^{-2}$. 
These corrections as well as the increase of uncertainties according to the aforementioned prescriptions, are taken into account in Table \ref{table:hyperfine}.\\

The spectroscopic quadrupole moments were calculated using the relation
\begin{equation}
Q_s^A = Q_s^{A_0} \frac{b_A}{b_{A_0}}
\end{equation}
with the reference values for $^{201}$Hg: \mbox{$Q_s^{A_0} = 0.387(6)$} b, $b_{A_0}=-280.107(5)$ MHz taken from \cite{Bieron2005_Quad201Hg_ref} and \cite{1961_Kohler}, respectively. 
The resulting spectroscopic quadrupole moments are shown in table \ref{table:hyperfine}.


\section{\label{sec:Discussion}Discussion}

\subsection{Changes in mean-square charge radii, \mbox{shape staggering} and `return to sphericity' of light mercury isotopes}

\begin{figure}[h]
\centering
\medskip
\includegraphics[width=1.0\columnwidth]{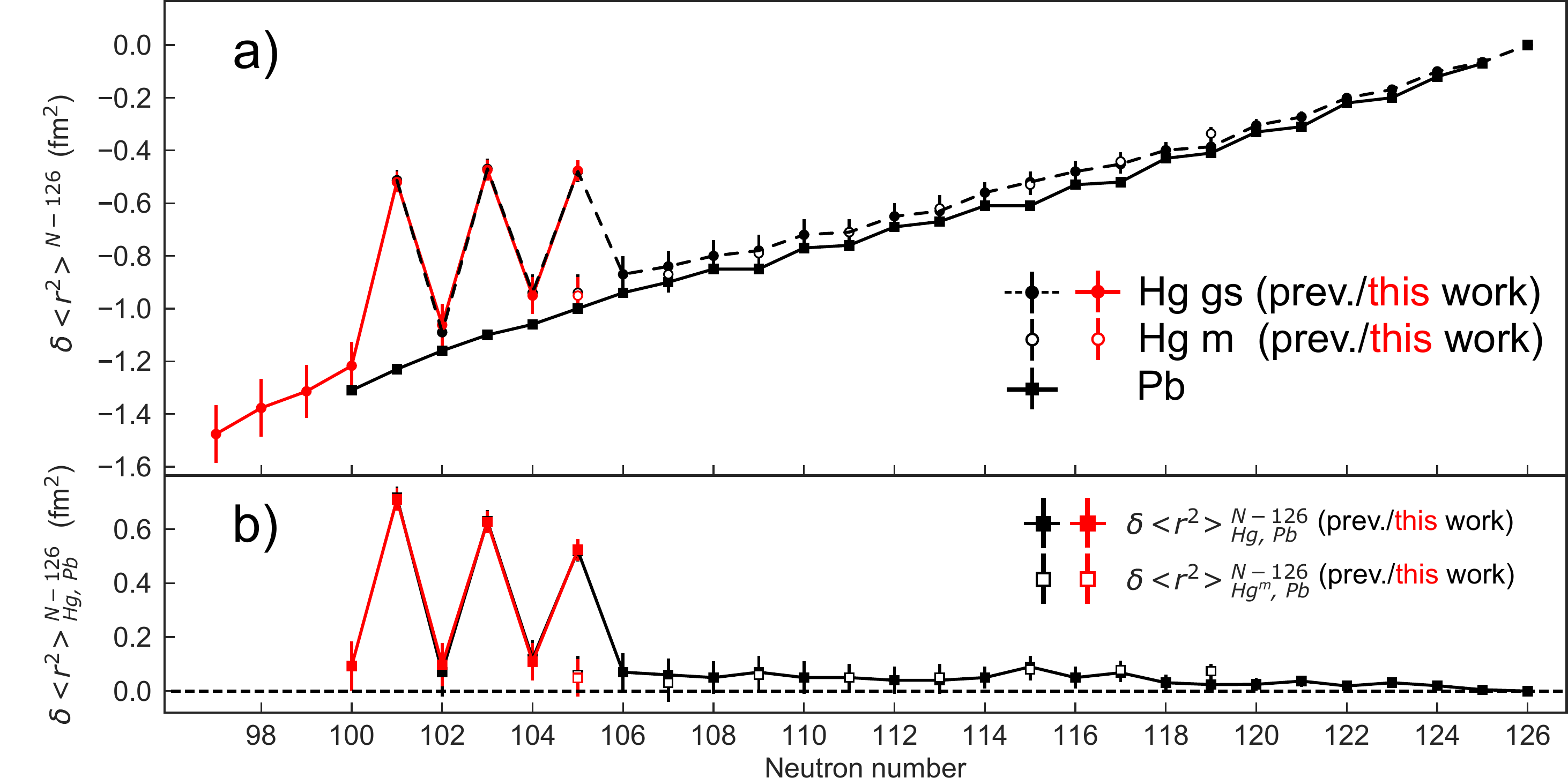}
\caption{a) Comparison of the change in mean-square charge radius for mercury and lead isotopic chains as a function of neutron number $N$, using $N=126$ as a reference. \\
b) Difference between the mercury and lead changes in charge radii where $\delta \langle r^2 \rangle ^{N-126}_{Hg,Pb}$ is defined as $\delta \langle r^2 \rangle ^{N-126}_{Hg} - \delta \langle r^2 \rangle ^{N-126}_{Pb}$. Data points shown in red and black correspond to this work and previous work respectively (\cite{Marinova2013} for lead, \cite{Ulm1986} for mercury).}
\label{fig:Charge_radii_comp}
\end{figure}

The change in the nuclear mean-square charge radius, \mbox{$\delta \langle r^2\ \rangle $}, for lead 
\cite{ANSELMENT1986,Dutta1991,DeWitte2007,Seliverstov2009,Marinova2013} and mercury (\cite{Ulm1986} and this work) isotopes are plotted with respect to $N=126$ at the top of Fig. \ref{fig:Charge_radii_comp}.
Three distinctly different regions are observed in the mercury charge radii. 
Mercury isotopes with $N>105$ follow a smooth trend, identical to the one of the isotopic chain of lead.
At $100<N<106$, in the neutron mid-shell region between the closed shells at $N=82$ and $N=126$, a large shape staggering is observed. 
Here, ground-state radii of the odd-$A$ mercury isotopes deviate substantially from the trend of the lead isotopes, 
which were found to keep their near-spherical shape at and beyond the neutron midshell \cite{DeWitte2007,Seliverstov2009}.
See also Fig. \ref{fig:Charge_radii_comp} b), where the difference in \mbox{$\delta \langle r^2\ \rangle $} between mercury and lead isotones is shown.
From the data obtained in the present work, it is observed that the staggering stops at $^{180}$Hg and \mbox{$\delta \langle r^2\ \rangle $} for mercury isotopes returns to the trend of lead.
In-beam recoil-decay tagging measurements, 
\cite{Carpenter1997,Kondev2000,Kondev2001,2001_Julin_intruderstates_RDT,2002_Kondev_PLB_179Hg_RDT,2003_melerangi_PRC_177Hg_RDT}, 
showed that the band-head energies of strongly-prolate deformed intruder bands in even-$A$ isotopes increase rapidly for mercury isotopes with decreasing neutron number at $N<101$.
The in-beam studies by Kondev {\it et al.} \cite{Kondev2000} and $^{181}$Pb $\alpha$-decay analysis by Jenkins {\it et al.} \cite{2002_jenkins_RDT_179Hg} have shown that a pronounced structural change takes place when moving from $^{181, 183}$Hg to $^{179}$Hg. 
Based on their decay scheme deduced in \cite{Kondev2000}, the authors proposed that the ground state of $^{179}$Hg is near-spherical with a possible weak-prolate deformation, 
rather than a strong, prolate deformation.
A similar interpretation was proposed for lighter odd-$A$ mercury isotopes with $A = 173-177$ 
\cite{2003_melerangi_PRC_177Hg_RDT,2009Odonnel,ODonnel2012}.
While those different studies had shown some indications of the shape of the ground state,
our data provide a direct measurement of the ground-state charge-radii changes.
The return of the lightest mercury nuclei to the trend of the weakly deformed mercury isotopes with N$>$105 (and to the near-spherical trend in lead nuclei) 
delineates the region of shape staggering to near the neutron mid-shell region at \mbox{$100<N<106$}.

\onecolumngrid
\clearpage

\twocolumngrid

\subsection{\label{sec:moments}Magnetic moments and spins of $^{177,179}$Hg}

The ground-state spin and parity of $^{179}$Hg was previously assigned \mbox{$I^{\pi}$ = (7/2$^-$)}, based on experimental data obtained for the $\alpha$-decay of $^{183}$Pb and subsequent $\alpha$-decays to daughter ($^{175}$Pt) and grand-daughter ($^{171}$Os) nuclei \cite{2002_Kondev_PLB_179Hg_RDT,2002_jenkins_RDT_179Hg}.
The same assignment, $I^{\pi}$ = (7/2$^-$), was proposed for the ground state of $^{177}$Hg, based on decay properties of the 13/2$^+$ isomeric state in this nucleus \cite{2003_melerangi_PRC_177Hg_RDT} and $\alpha$-decay of $^{181}$Pb \cite{Andreyev2009}. 
As was indicated in Sec. \ref{sec:Fitting}, we also tested $I=9/2$ as a possible assignment,
since the $\nu f_{7/2}$ and $\nu h_{9/2}$ orbitals are assumed to play a dominant role in the negative-parity states around $N=97,99$.
The quality of fitting is the same for both assumptions.
However, the measured large and negative magnetic moments of $^{177,179}$Hg rule out a $\nu h_{9/2}$ hole configuration as this is expected to give a large positive magnetic moment \mbox{$\mu_{\text{th}}$($\nu h_{9/2}$) = +0.69 $\mu_N$} \cite{Bauer1973}.\\

Two plausible origins of the $^{177,179}$Hg spins and magnetic moments are considered:\\
a) They arrise from the strong Coriolis mixing of the Nilsson states of $\nu f_{7/2}$ and $\nu h_{9/2}$ parentage at very low deformation ($\beta_2 \approx$ 0.05 - 0.07) or \\
b) they may be regarded as spherical $\nu f_{7/2}$-hole states. 

Let us first focus on the explanation a).\\
In \cite{2002_Kondev_PLB_179Hg_RDT}, two likely candidates for the ground-state Nilsson configuration of $^{179}$Hg were proposed: the 7/2$^{-}$[514] and 7/2$^-$[503] orbitals, arising from the $\nu h_{9/2}$ and $\nu f_{7/2}$ neutron shell-model states respectively. 
These orbitals come close to the Fermi level only for small prolate deformations \mbox{($\beta_2$ $<$ 0.15)}. 
In $^{175}$Pt, the $\alpha$-decay daughter of $^{179}$Hg, the $\nu h_{9/2}$ 7/2$^{-}$[514] orbital was chosen as preferable for the ground state band \cite{2003_melerangi_PRC_177Hg_RDT}.
However, several measured magnetic moments of the 7/2$^{-}$[514] Nilsson state of deformed $N$ = 105 nuclei ($^{183}$Pt$^{m}$,$^{177}$Hf,$^{175}$Yb)
are close to $\mu \approx +0.8 \mu_N$ \cite{Stone}. 
Nilsson-model calculations describe these experimental data fairly well \cite{Ekstrom1976}.
The calculations with the same approach predict positive moments for $^{177,179}$Hg even at rather low deformation 
(\mbox{$\mu$(7/2$^{-}$[514], $^{179}$Hg)$_{\text{th}}$ = +0.46 $\mu_N$ and +0.34 $\mu_N$} at \mbox{$\beta_2$ = 0.15 and 0.10} respectively).
Coriolis-mixing might however play a role as the lowest states in the lightest mercury isotopes display a high spin at low deformation. 
At \mbox{$\beta_2$ $<$ 0.15}, contributions of the different Nilsson-orbitals stemming from the $\nu f_{7/2}$ and $\nu h_{9/2}$ orbitals 
to the lowest 7/2$^-$ state become nearly equal. This mixing would bring the magnetic moment down in value with respect to a pure 7/2$^{-}$[514] configuration and might be a possible explanation of the measured magnetic moments of $^{177,179}$Hg.

\begin{figure}[h]
\centering
\includegraphics[width=1.0\columnwidth]{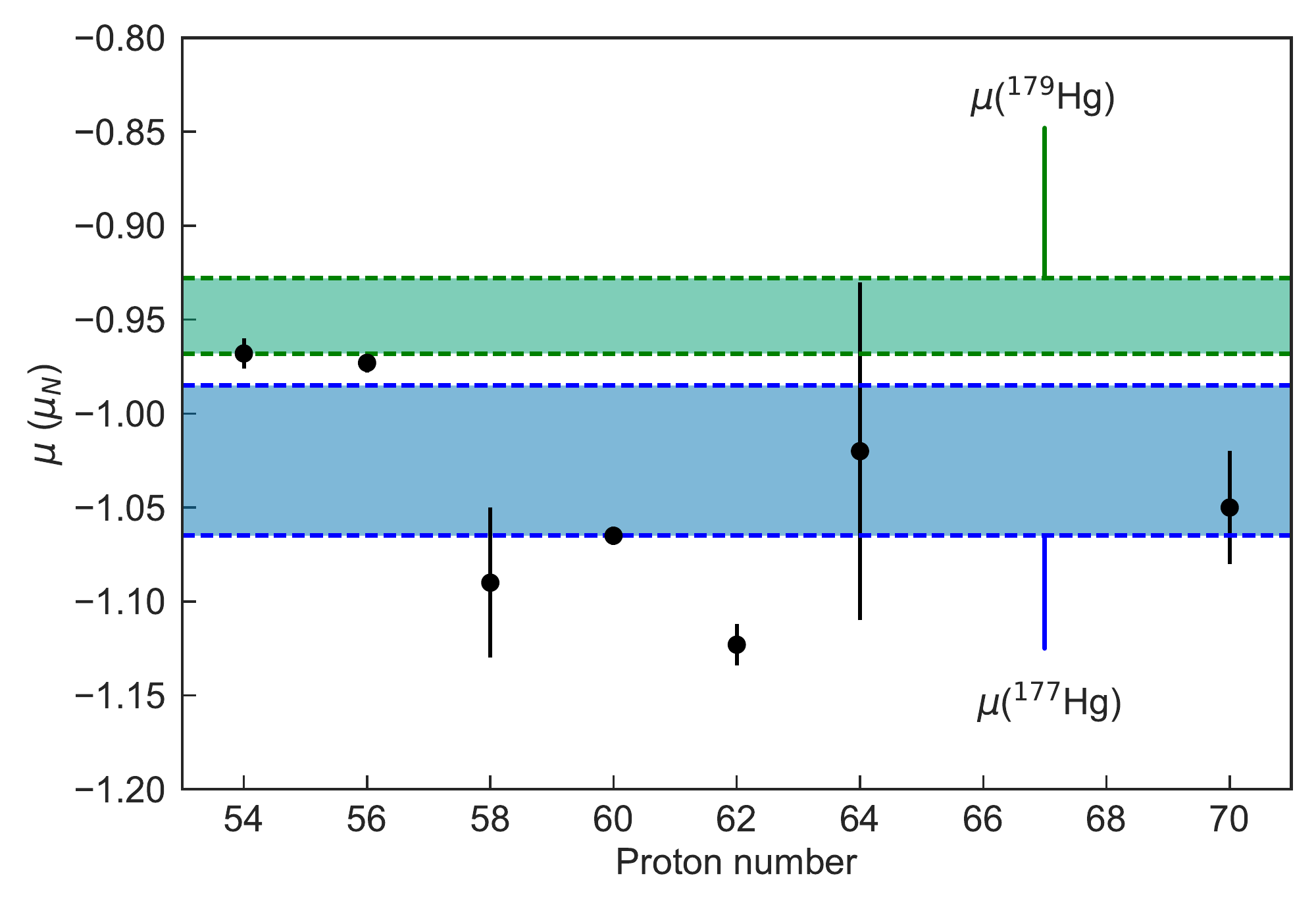}
\caption{Comparison of magnetic moments for $N=83$ isotones, having one neutron in the $(\nu f_{7/2})$ shell above the closed shell $N=82$,
with the measured $\mu(^{177}\text{Hg})$ and $\mu(^{179}\text{Hg}$) indicated by shaded boxes. Data from \cite{Stone}, \cite{2002Barzakh} and references therein.}
\label{Fig:sp_mu_comparison}
\end{figure}

Let us now discuss option b) and explore the interpretation of the ground-states of $^{177,179}$Hg as spherical $\nu f_{7/2}$-hole states.\\

Magnetic moments of $^{177, 179}$Hg come comparatively close to the single particle estimation for spherical neutron shell $\nu f_{7/2}$: \mbox{$\mu_{\text{s.p.}}$ = -1.3 $\mu_N$} (with the commonly adopted renormalization of the neutron $g$ factors: \mbox{g$_\text{s}^{\text{eff}}$ = 0.6g$_\text{s}^{\text{free}}$} and g$_\text{l}^{\text{eff}}$ = -0.05).
It is instructive to compare the magnetic moments of the presumed 7/2$^-$ $^{177,179}$Hg ground states with measured magnetic moments of the ground states of $N=83$ isotones with one neutron in the $f_{7/2}$ shell.
This comparison is shown in Fig. \ref{Fig:sp_mu_comparison}.
One can see that \mbox{$\mu$($^{177, 179}$Hg)} corresponds to \mbox{$\mu$($\nu f_{7/2}$)} in the $N = 83$ isotones for which all show a rather large negative magnetic moment value.
If this interpretation is valid, then the ground states of $^{177,179}$Hg could be regarded 
as holes in the $\nu f_{7/2}$ orbital within a simple shell model picture. 

This means that for the light mercury isotopes, 
the state arising from a neutron hole in the $\nu f_{7/2}$ orbital lies above that arising from a neutron hole in the $\nu h_{9/2}$ orbital. 
As a consequence, the state ordering for $Z=80$ and $N<100$ is reversed with respect to the $N=83$ isotones in the vicinity of the stable isotopes, 
where the $\nu f_{7/2}$ orbital is filled first after the $N=82$ shell closure.
Surprisingly, with the increase of $Z$ after $_{80}$Hg, the normal ordering is restored in $_{81}$Tl and $_{82}$Pb.
The ground-state spin and parity of $^{181}$Pb$_{99}$ was determined as $9/2^-$ due to a hole in $\nu h_{9/2}$
shell arising from the complete depletion of the $i_{13/2}$ and $p_{3/2}$ shells lying above the \mbox{$N = 100$} spherical subshell closure \cite{Andreyev2009}. 
Similarly, the odd neutron state in $^{180}$Tl$_{99}$ was assumed to be a $\nu h_{9/2}$-hole state on the basis of its magnetic moment value \cite{Barzakh2017,Elseviers2011}. 
Thus, for Z $>$ 80 the $\nu h_{9/2}$ shell appears to be situated above the $\nu f_{7/2}$ shell. 

It was shown that the energy differences between the lowest-lying $9/2^−$ and $7/2^−$ states 
for the $N=83$ and $N=85$ isotones show a rapid drop above $Z=64$ (see \cite{Bianco2010} and references therein). 
According to Bianco {\it et al.} \cite{Bianco2010}, this drop reflects the gradual approach in energy of the $\nu h_{9/2}$ and $\nu f_{7/2}$ neutron single-particle orbitals. 
The presumed convergence of the $\nu h_{9/2}$ and $\nu f_{7/2}$ neutron levels also provides a natural explanation 
for the anomalous absence of charged-particle emission from the high-spin isomer of $^{160}$Re \cite{Darby2011}.
This shell evolution was explained in \cite{Bianco2010} by the influence of the tensor part of the nucleon-nucleon interaction \cite{Otsuka2006}. 
It was predicted in Ref. \cite{Bianco2010} that the energies of the neutron single-particle orbitals may become inverted for high $Z$. 
The interpretation of the $^{177,179}$Hg ground states as $\nu f_{7/2}$ shell-model states 
and their first excited states as predominantly $\nu h_{9/2}$ states \cite{2002_Kondev_PLB_179Hg_RDT,2003_melerangi_PRC_177Hg_RDT} is in agreement with this description.

\subsection{\label{sec:quadmoments}Quadrupole moments of $^{177,179}$Hg}

The quadrupole moments of $^{177,179}$Hg extracted from their respective HFS $b$ constants are consistent with a simple spherical shell model approach.
According to the seniority scheme \cite{Talmi1963}, quadrupole moments should be linearly dependent on the number of particles occupying a certain orbital as can be seen from equation \ref{eq.quad}.

\begin{equation}
\langle j^n | \hat{Q} | j^n \rangle = \frac{2j+1-2n}{2j+1-2\nu} Q_{\text{s.p.}} \label{eq.quad}
\end{equation}

\begin{equation}
Q_{\text{s.p.}} = -e \frac{2j-1}{2j+2} \langle r^2 \rangle_j \label{eq.quadscaling}
\end{equation}

Here, a $j^n$ configuration with $n$ nucleons is labelled by a seniority $\nu$, the number of unpaired neutrons ($\nu =1$ in our case).
The observed spectroscopic quadrupole moment $Q_s$ is represented by $\langle j^n | \hat{Q} | j^n \rangle$, while $Q_{\text{s.p.}}$ indicates the single-particle quadrupole moment value.
Recently, the seniority scheme with a linear dependence of $Q_s$ on the number of particles in the filling of a shell has been found to work remarkably well for the cadmium, astatine and actinium isotopes \cite{Yordanov2013,Cubiss2018,Ferrer2017}.
One can check the validity of the seniority scheme for the filling the $\nu f_{7/2}$ orbital with the assumption that $^{177,179}$Hg have 5 and 7 neutrons in this shell $(n=5,7)$, respectively. For $n=1,3,5$ we choose $^{145,147,149}$Sm \cite{Stone}. The values of $Q$($^{177,179}$Hg) were scaled according to Eq. \ref{eq.quadscaling} by 16\% using the global evaluation of $\langle r^2\ \rangle $ data in \cite{Marinova2013}, to remove the dependence of $Q$ on $\langle r^2\ \rangle $, which distorts the presumed linear dependency. 

\begin{figure}[h]
\centering
\includegraphics[width=1.0\columnwidth]{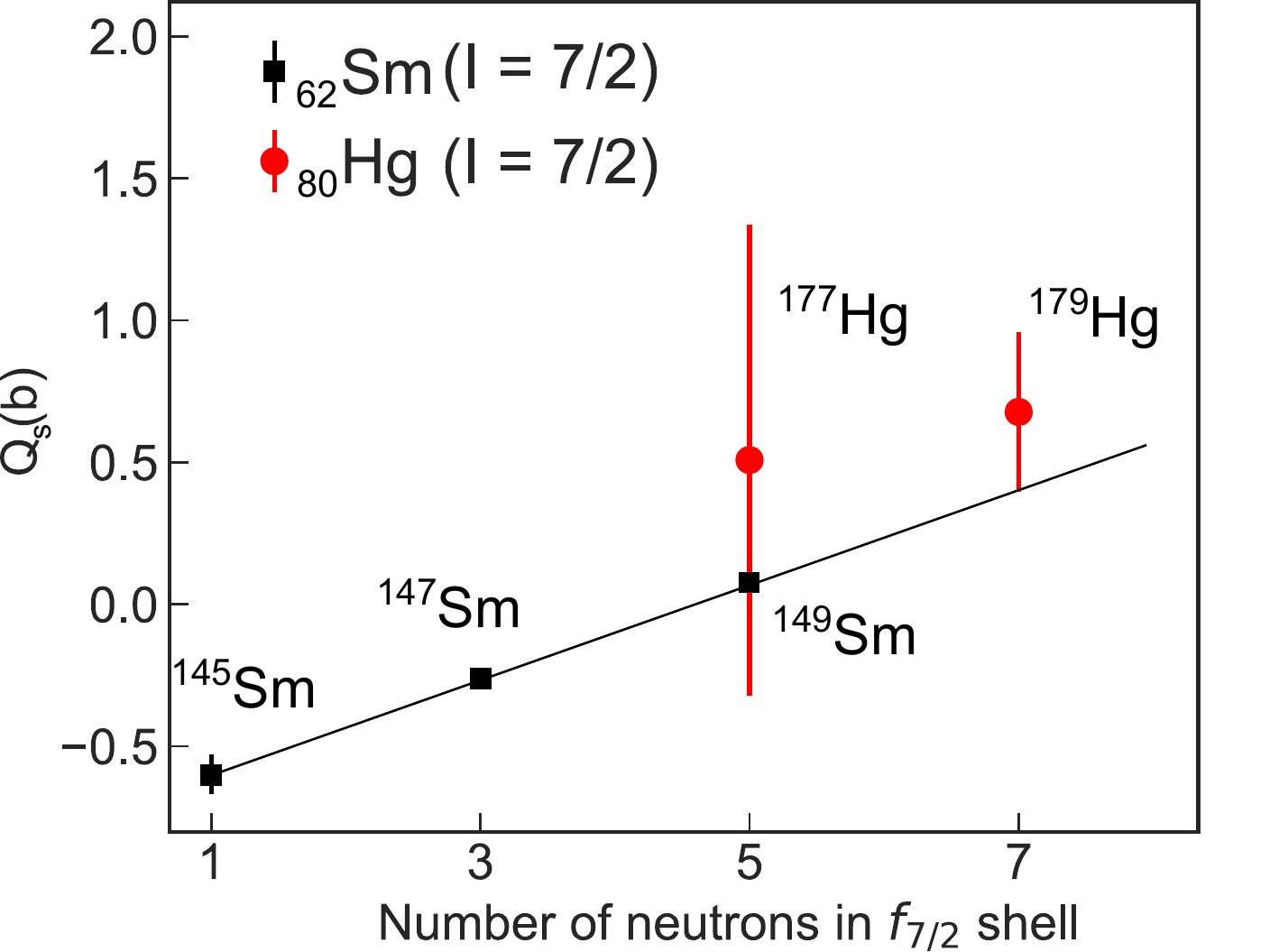}
\caption{Comparison of quadrupole moments for the $^{145,147,149}$Sm isotopes (data from \cite{Stone}), with measured values for $^{177,179}$Hg corrected for the difference in charge radius between samarium and mercury.}
\label{Fig:seniority_Q_comparison}
\end{figure}

As can be seen from Fig. \ref{Fig:seniority_Q_comparison}, the light mercury isotopes follow the trend given by samarium even though the considered nuclei have very different neutron numbers: $N=83-87$ for samarium and $N=97-99$ for mercury.
This observed correspondence of the $Q_s$($^{177,179}$Hg) values to the seniority-scheme prediction supports the assumption of the $\nu f_{7/2}$-hole nature of the $^{177,179}$Hg ground states.

\subsection{Comparison with Nuclear Density Functional Theory (DFT) and Monte-Carlo Shell Model (MCSM) calculations}

Frauendorf and Pashkevich \cite{Frauendorf1975} described shape staggering in the mercury charge radii 
using a microscopic-macroscopic approach with Strutinsky shell corrections, 
thereby proving that mean-field models are able to predict the shape staggering and shape coexistence in mercury.
More recently, the even-even mercury isotopes were studied and spectroscopic observables (e.g. charge-radii) were calculated
using an Interacting Boson Model with configuration mixing (IBM-CM) \cite{GarciaRamosHeyde2014} and a beyond mean field approach \cite{Yao2013}.
In both approaches, the deformation energy surfaces for nuclei near the midshell have a lowest-energy minimum at prolate deformation, 
accompanied by a second, oblate minimum (for $176<A<186$ in \cite{Yao2013} and $180<A<186$ in \cite{GarciaRamosHeyde2014}).
This is in contrast to experimental results, that indicated that the even-even mercury isotopes have weakly deformed (oblate) ground-state shapes \cite{Bree2014}.
Both papers \cite{Yao2013,GarciaRamosHeyde2014} only discuss the even-$A$ mercury cases 
and do not go deeper into the origin of the experimentally observed shape staggering in the odd-$A$ mercury isotopes or the prediction of its magnitude.\\

Early DFT calculations for both even- and odd-mass isotopes mercury isotopes with 
the SLy4 parameterization achieved a reproduction of the odd-even radius 
staggering for $N>100$, by fitting the pairing strength to the one-particle 
separation energy of these isotopes~\cite{Sakakihara03}. 
In ~\cite{(Sar15)} the shape coexistence in the region was confirmed as a mechanism for the observed radius
staggering for the SLy4, Sk3 and SGII parameterizations with fixed gap 
parameters, although the exact staggering could not be reproduced. 
In \cite{(Mir17)}, the SLy4 parameterization was again employed, 
but this time with a pairing strength adjusted to the odd-even staggering of the lead isotopes, 
leading to a qualitative correct staggering that is off by four neutron numbers. 
As explicitly noted in \cite{(Mir17)}, the  precise staggering pattern depends 
sensitively on the details of the effective interaction. This can be explicitly 
demonstrated by comparing results for the SLy4 parameterization from all
three sources, \cite{Sakakihara03,(Sar15)} and \cite{(Mir17)}, which differ 
in the pairing interaction employed in the respective calculations.
Altogether, the past DFT investigations led to an overall understanding of the physics of the phenomenon.
However, because of the known deficiencies of the available parameterisations of 
the functional, achieving a quantitative description of all its details is still not possible.\\

In an attempt to understand the behavior exhibited by the light mercury isotopes, 
and to extend the description to odd-$A$ mercury isotopes,
Density Functional Theory (DFT) and large-scale Monte-Carlo Shell model (MCSM) calculations were performed in this work
which are described in sections \ref{sec:DFT} and \ref{sec:MCSMcalculations} respectively.


\subsubsection{\label{sec:DFT} \bf{DFT calculations}}

In a mean-field picture, a dramatic staggering of the isotope shift as observed
in experiment can be achieved as a consequence of a ground-state \emph{shape} staggering. 
In order for a shape staggering to occur and produce a large change
in nuclear charge radius, three conditions need to be met. 
First, the isotopes must have multiple competing shapes, 
which in the case of neutron-deficient mercury isotopes, 
are oblate, weakly prolate and strongly prolate. 
Second, the minima must exhibit sufficiently different deformations, 
leading to substantial differences in the corresponding calculated root-mean square radius. 
For the mercury isotopes, the even-$A$ isotopes should have weakly-deformed minima, 
while the odd-$A$ nuclei in the region should have strongly-prolate deformations. 
Third, the excitation energy of the strongly-prolate minimum should be small for
a specific set of nucleon numbers, $101 \leq N \leq 105$. 
More precisely, it should be comparable to the odd-even \emph{mass} staggering for these isotopes i.e the difference in binding
energy between odd-$A$ isotopes and their even-$A$ neighbours.

In this way, the difference in odd-even mass staggering in both wells can shift 
the energetic balance between the weakly- and strongly-deformed configurations. 
As is argued below, these conditions put extremely stringent constraints on the parameters of the DFT 
functionals, far beyond the precision with which these parameters 
have been determined so far. 


Two sets of systematic calculations for the mercury isotopes were performed. 
The first set considers axially-symmetric configurations obtained using the HFBTHO code~\cite{(Sto05a),(Sto13)}, 
for six different Skyrme functionals UNEDF0 \cite{UNEDF0}, UNEDF1 \cite{UNEDF1}, UNEDF1$^\text{SO}$
\cite{UNEDF1_2}, SLy4 \cite{Chabanat1998}, SkM* \cite{Bartel1982} and SGII \cite{Vangiai1981}. 

The second set is comprised of 3D coordinate-space calculations with the MOCCa code~\cite{RyssensPhD}, 
employing the eight parameterizations of the recent SLy5sX~\cite{Jodon16} family 
supplemented by a zero-range surface pairing interaction~\cite{Krieger1990} 
with a standard pairing strength~\cite{Rigollet1999}. 
In both sets of calculations, we retained the blocked configurations for the odd-$A$ isotopes 
that are lowest in energy, which in general do not correspond to the experimental ground state spins and parities.

None of these fourteen parameterizations predict potential energy surfaces (PES) 
that fulfill all three conditions referred to above. 
However, all of them at least predict near-degenerate oblate and prolate minima to coexist in light mercury isotopes.
This feature thus seems to be a generic property of the bulk macroscopic energy, 
and is independent of the detailed orderings of single-particle levels 
that differ between functional parameterizations. 
In particular, the orbitals corresponding to the experimental 
ground-state spins and parities do not appear at the Fermi surfaces of odd-$A$ mercury isotopes.

We will focus in what follows on four different parameterizations. 
The first, UNEDF1$^\text{SO}$, is based on the UNEDF1 parameterization 
which has been adjusted to global nuclear properties across the nuclear chart, 
but with modified spin-orbit and pairing properties to reproduce detailed spectroscopic 
properties of nuclei around $^{254}$No. 
Calculations here have been carried out for three different variants: 
one employing the pairing prescription from \cite{UNEDF1_2}, 
and two others with an amplified pairing strength compared to 
the original parameterization, by 8\% and 20\% respectively.

The three remaining functionals, SLy5s1, SLy5s4 and SLy5s8, are members of 
a family of parameterizations constructed on the basis of SLy5*~\cite{Pastore13} 
to study the impact of varying the surface tension of Skyrme functionals. 
SLy5s8 has a surface tension similar to SLy4 and SLy5*, while for SLy5s7 down to SLy5s1, 
it takes progressively smaller values. 
Because of the nature of the fitting protocol, all members of the family exhibit similar single-particle structures. 
However, the deformation properties of the parameterizations are quite different, 
which is a consequence of the differences in surface tension. 
In particular, the parameterization with lowest surface tension, SLy5s1,
gives quite a satisfying description of the fission barriers of heavy nuclei, 
such as actinides and $^{180}$Hg~\cite{Ryssens18}. 
Here, results are presented for SLy5s1, SLy5s4 and SLy5s8 representing 
functionals with low, intermediate and high surface tension respectively.

\begin{figure}
\centering
\includegraphics[width=1.0\columnwidth]{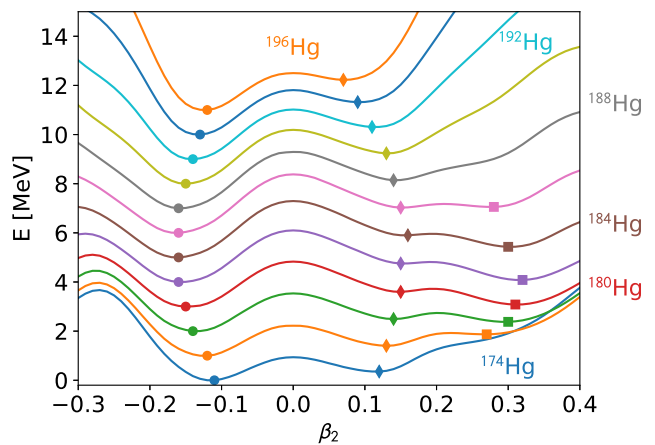}
\caption{Potential energy surfaces calculated using the 
UNEDF1$^\text{SO}$ functional for even-$A$ mercury isotopes between \mbox{$A=174$} and
\mbox{$A=196$} as a function of the axial quadrupole moment, $\beta_{2}$. Energies are
renormalized to the minimum of the curve, with an additional offset of 1 MeV 
between isotopes. Circles indicate the oblate minima, diamonds the weakly
prolate minima, and squares the strongly-prolate minima (if present).}
\label{Fig:UNIDEF_PES}
\end{figure}

The PES for the even-$A$ neutron-deficient mercury 
isotopes as calculated with UNEDF1$^\text{SO}$ with standard pairing 
as a function of the axial quadrupole deformation, are shown in Fig.~\ref{Fig:UNIDEF_PES}. 
For all isotopes, the lowest minimum is the oblate one. 
In a narrow region, $86 \leq N \leq 106$, the strongly-prolate minimum is very close in energy 
to both the strongly-prolate and the oblate minimum. 
In this way, the UNEDF1$^\text{SO}$ PES fulfills the first two conditions for producing a sharp radius staggering. 
The other parameterizations discussed here give qualitatively similar energy curves with the same overall pattern of minima, 
but with slightly different relative energy between them. 
Albeit small, these differences can change the energetic order of minima when going from one parameterization to the next.

\begin{figure}
\centering
\includegraphics[width=1.0\columnwidth]{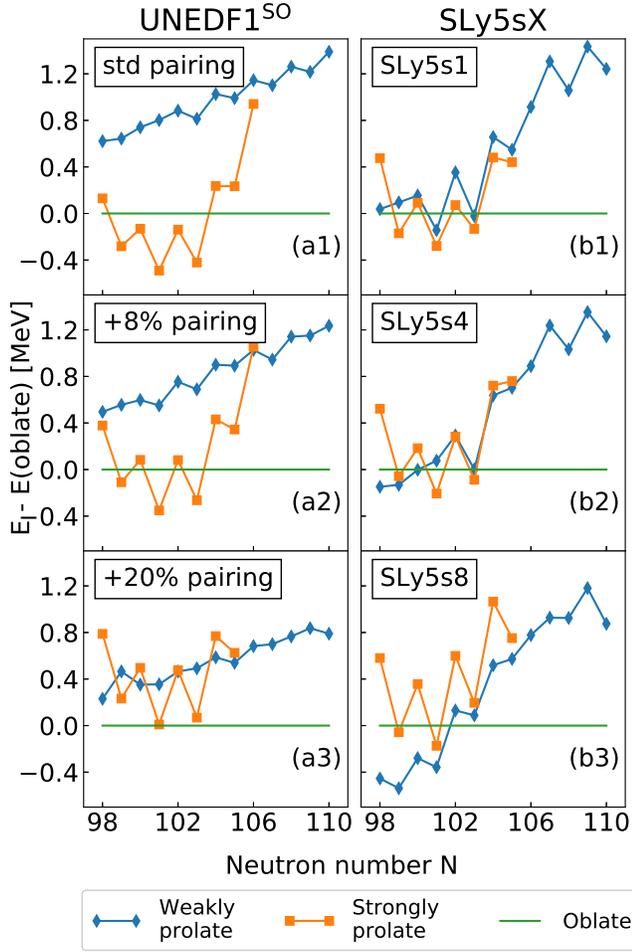}
\caption{Differences between energies of weakly and strongly
prolate configurations and those corresponding to the oblate configurations, 
as calculated with UNEDF1$^\text{SO}$ (left) and the SLy5sX parameterizations
(right). For UNEDF1$^\text{SO}$ calculations with standard pairing (top),
pairing amplified by 8\% (middle) and pairing amplified by 20\% (bottom) are
presented. For the SLy5sX family, results for SLy5s1 (top), SLy5s4 (middle)
and SLy5s8 (bottom) are presented.}
\label{Fig:UNIDEF_DIF}
\end{figure}

To appreciate the detailed energy balance between the different configurations in both odd- and even-$A$ isotopes, 
the excitation energies of the weakly- and strongly-prolate minima are shown in Fig.~\ref{Fig:UNIDEF_DIF}
with respect to the oblate minimum for UNEDF1$^\text{SO}$ with different values of the pairing strength, 
as well as for SLy5s1, SLy5s4 and SLy5s8. 
An unaltered UNEDF1$^\text{SO}$, Fig.~\ref{Fig:UNIDEF_DIF}, 
predicts all isotopes between $99 \leq N\leq 103$ to be strongly-prolate deformed. 
Increasing the pairing strength by 8\% suffices to change the predicted deformation of the even-$A$ isotopes in that region, 
producing a staggering pattern.
Further increasing the pairing strength makes the shape staggering completely disappear: 
the oblate minimum is the lowest for all isotopes. 
For SLy5s1 as well, the ground state staggers between strongly-prolate and oblate minima in the region $99 \leq N \leq 103$.
By increasing the surface tension the shape staggering can be changed: 
for SLy5s4 only two odd isotopes exhibit strong prolate deformation while none do for SLy5s8.

For the SLy5sX parameterizations, larger surface tension penalizes the strongly-prolate minimum 
compared to the oblate and weakly-prolate ones because of its larger deformation. 
Increasing the pairing strength serves a similar purpose: 
the strongly-prolate minimum gains in energy compared to the oblate minimum. 
Note that for all variants of UNEDF1$^\text{SO}$ the size of the odd-even mass staggering in the strongly-prolate minimum
is almost independent of the size of the pairing strength: 
the mass staggering is at the root of the shape staggering, 
but the detailed fine tuning of the staggering is achieved by changing the balance of the minima 
and not through a change of the size of the odd-even mass staggering. 
Although being quite different features of the effective interaction, 
the variation of the pairing strength and the surface tension can both be employed 
to change the balance of minima in order to fine-tune the shape staggering.

\begin{figure}
\centering
\includegraphics[width=1.0\linewidth]{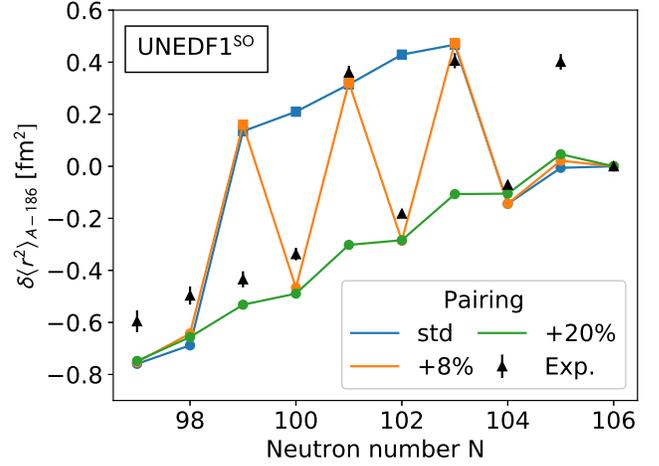}
\caption{Comparison of the calculated isotopic shift using the 
         UNEDF1$^\text{SO}$ paramaterization for three different values of the 
         pairing strength with experimental data, normalized to 
         $^{186}$Hg. As in Fig. \ref{Fig:UNIDEF_PES} and Fig. \ref{Fig:UNIDEF_DIF}, circles indicate oblate, diamonds weakly-
         prolate and squares strongly-prolate configurations.}
\label{Fig:UNIDEF_vs_exp}
\end{figure}

\begin{figure}
\centering
\includegraphics[width=1.0\linewidth]{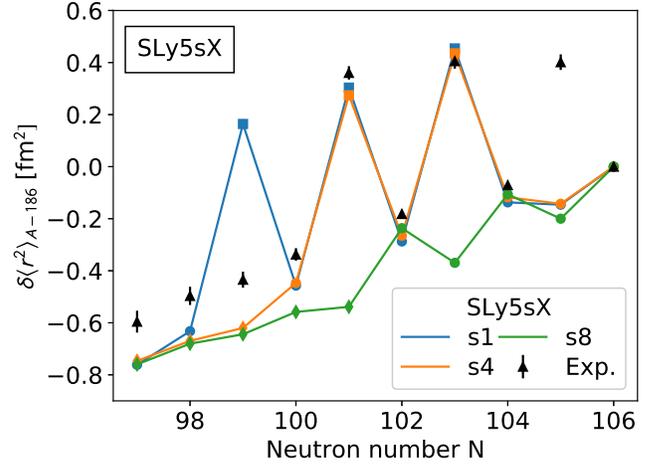}
\caption{Same as Fig.~\ref{Fig:UNIDEF_vs_exp}, but for the SLy5s1, SLy5s4 and 
        SLy5s8 parameterizations. Circles, diamonds and squares indicate oblate, weakly-
         prolate and strongly-prolate configurations, respectively.}
\label{Fig:SLy5sX_vs_exp}
\end{figure}

The effect of the shape staggering on the calculated radius staggering 
is shown in Fig.~\ref{Fig:UNIDEF_vs_exp} for the set of UNEDF1$^\text{SO}$
calculations and in Fig.~\ref{Fig:SLy5sX_vs_exp} for the SLy5sX calculations. 
Large charge radii appear where the calculated ground state has a strong, prolate deformation. 
SLy5s1 and UNEDF1$^\text{SO}$ with 8\% increase in pairing strength 
predict a staggering pattern that sets in later with reducing neutron number. 


The presented DFT results show that the origin of the radius staggering 
can be properly identified in terms of the ground-state shape staggering between the oblate 
(or strongly-prolate) and strongly-prolate configurations along the chain of mercury isotopes. 
The effect is extremely sensitive to the fine details of the functional: 
by slightly varying the pairing strength of (UNEDF1$^\text{SO}$) 
or the surface tension (SLy5SX) of the functional, 
one can fine tune the detailed balance between the three, coexisting minima, 
which can change dramatically the shape staggering and hence the radius staggering.

While this effect is demonstrated here for the pairing strength and surface tension, 
the balance between the minima also depends on many other properties of the functionals 
both in the particle-hole and particle-particle channels. 
Because of this sensitivity, it is very unlikely that any of the existing parameterizations 
can be tuned to reproduce exactly this very particular shape-coexistence pattern. 
For this reason, we can not meaningfully state a strong preference for any parameterization discussed here. 
The precise staggering pattern could, however, be used to impose stringent conditions on future parameter fits.

\subsubsection{\label{sec:MCSMcalculations} \bf{MCSM calculations}}
Monte Carlo Shell Model (MCSM) calculations are a type of configuration-interaction approach 
for atomic nuclei that uses the advantages of quantum Monte-Carlo, variational and matrix-diagonalization methods \cite{Shimizu2017}. 
This is the first time such calculations were performed for such a heavy system as the mercury isotopes
and they are the heaviest MCSM calculations so far.
The massively parallel K-supercomputer \cite{Kcomp} provided the computing power to execute the calculations.

The model-space single-particle orbitals used in these calculations consist of proton orbitals from 1$g_{7/2}$ up to 1$i_{13/2}$ and neutron orbitals from 1$h_{9/2}$ up to 1$j_{15/2}$,
using the doubly-magic $^{132}$Sn nucleus as inert core. 
As a result, a large number of nucleons (30 protons and up to 24 neutrons) were left to interact in a large model space.

In these orbitals, all nucleons interact through effective nucleon-nucleon ($NN$) interactions.
The neutron-neutron ($n-n$) and proton-proton ($p-p$) interactions are taken from \cite{BABrownPb}, while the proton-neutron ($p-n$) interaction from \cite{vmu} was used.  
Effective charges for protons and neutrons \cite{Qbeta} being 1.6$e$ and 0.6$e$ were used together with a spin-quenching factor of 0.9 \cite{Qbeta} and
single-particle energies were adjusted to properties of doubly-magic $^{132}$Sn and $^{208}$Pb nuclei.


Eigenstates were calculated for mercury isotopes with $177 \leq A \leq 186$ with excitation energies below \mbox{2 MeV} 
and spins and parities as observed in experiment.
For each of these eigenstates, the magnetic moment, quadrupole moment, excitation energy and nucleon occupation numbers were computed.

\begin{figure}[h]
\centering
\includegraphics[width=1.0\columnwidth]{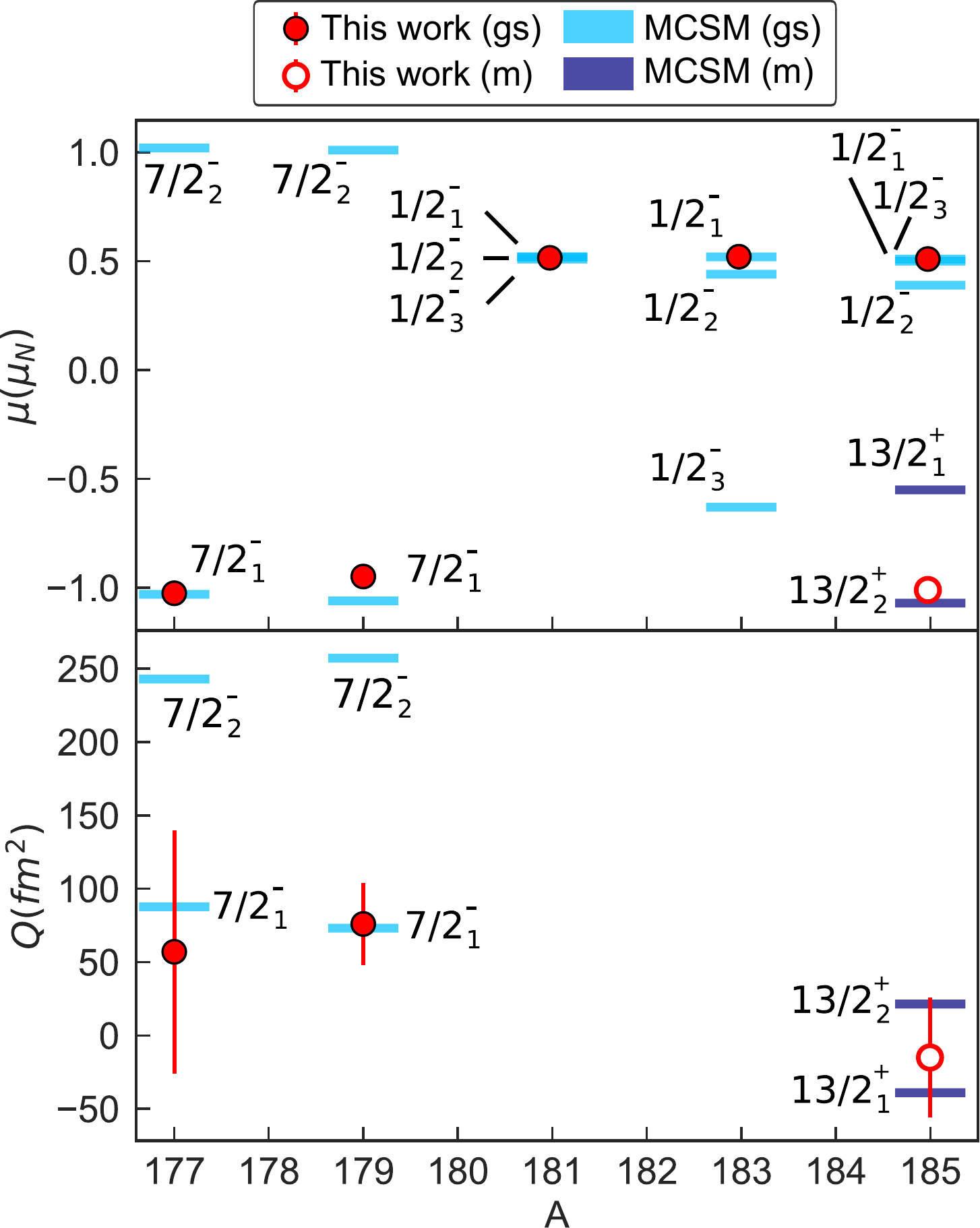}
\caption{Comparison of experimental magnetic moment values from laser spectroscopy with MCSM calculations for odd-$A$ neutron-deficient mercury isotopes ($A=177-185$).
Red circles refer to experimental data. }
\label{Fig:moments}
\end{figure}

As shown in Fig. \ref{Fig:moments}, the MCSM calculations reproduce the magnetic and quadrupole moments for both experimentally-observed ground states 
and isomeric states of all measured odd-$A$ isotopes. 
While for some isotopes the electromagnetic moment data clearly favor one eigenstate over the others, such a distinction cannot be made for all isotopes.
The magnetic moment of $^{177}$Hg for instance clearly favors the $7/2^-_1$ state over $7/2^-_2$, while in $^{181}$Hg, 
the magnetic moment of all $1/2^-$ states is nearly identical. MCSM calculations were also performed for $9/2^-$ states in $^{177,179}$Hg. The magnetic moments of these states were found to be large and positive: $\mu(9/2^-) = 1.26, 0.72$ and $1.26, 1.47$ $\mu_N$ for the first two excited states with $I^{\pi} = 9/2^-$ in $^{177,179}$Hg, respectively.\\

The MCSM eigenstate is given by a superposition of MCSM basis vectors. Each MCSM basis vector is a deformed Slater determinant, 
for which intrinsic quadrupole momenta $Q_0$ and $Q_2$ can be calculated. $Q_0$ and $Q_2$ can be used as “partial coordinates”, 
from which the nuclear shape parameters $\beta_2$ and $\gamma$ can also be extracted by standard relations \cite{bohr1998nuclear}. 
A given MCSM basis vector is placed as a circle on the Potential Energy Surface (PES) according to $Q_0$ and $Q_2$. 
The importance of this basis vector in the eigenstate is expressed by the area of a circle, being proportional to the square of the overlap to the eigenstate. 
This is called the T-plot \cite{Tsunoda}.

The calculated $\beta_2$ values are extracted via the following expression:
\begin{equation}
	Q_0 = \frac{3Z}{\sqrt{5 \pi}} R^2 \langle \beta_2\ \rangle (1+0.36\langle \beta_2\ \rangle ) \label{eq:Q0_to_beta2}
\end{equation}
where \mbox{$R^2$} is calculated as \mbox{$R^2 = (1.2 \text{ fm}$ $A^{1/3})^2$} \cite{Otten1989}.
Since the mass quadrupole moment, rather than the electric quadrupole moment results from MCSM calculations, the factor $Z$ was replaced in Eq. \ref{eq:Q0_to_beta2} by $A$.
The correction term $0.36 \langle \beta_2\ \rangle $ was neglected, because $\langle \beta_2\ \rangle $ is at most about 0.2 presently.
The resulting equation becomes
\begin{equation} 
\langle \beta_2\ \rangle  = Q_0 \left( \frac{\sqrt{5 \pi}}{3 A R^2} \right) \label{eq:beta2_from_Q0}
\end{equation}
corresponding to the equations used in \cite{Utsuno2015,Rodriguez2010}.

\begin{figure}[ht]
\begin{center}
\includegraphics[width=\columnwidth]{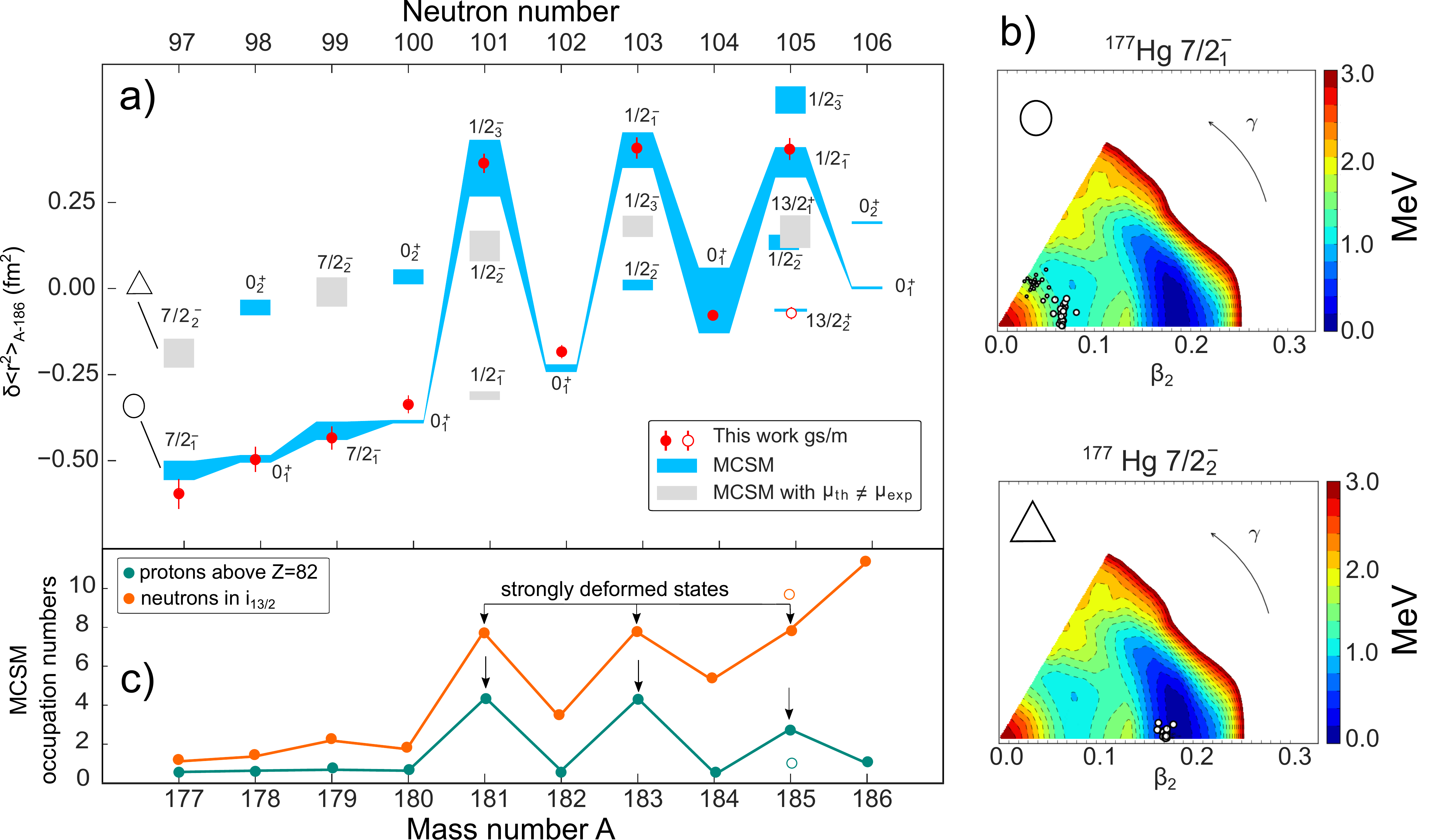}
\end{center}
\caption{ {\bf  a) }: Mean square charge radius relative to that of the ground state of
$^{186}$Hg. Red points are experimental data from this experiment.
The blue shaded area indicates the radii corresponding to MCSM eigenstates
for the observed ground states, $J^{\pi}$ = 0$^+$ for even-$A$, 1/2$^-$ for
$^{181,183,185}$Hg, and 7/2$^-$ for $^{177,179}$Hg.  
In addition, the \mbox{$\delta\langle r^2\ \rangle $} value of the isomeric state 13/2$_2^+$ is shown by the lower line for $^{185}$Hg. 
The width of the blue shaded areas corresponds to the spread of MCSM basis-vector deformation parameters on the PES.  
The gray areas show MCSM eigenstates for which the calculated magnetic moments differ from measured values.
{\bf b)}: T-plots for two states in $^{177}$Hg are shown on the right (see text for more information).
{\bf c)}: Proton and neutron occupation numbers.
}
\label{Fig:charge_radius}
\end{figure}

The differences in mean-square charge radii \mbox{$\delta\langle r^2\ \rangle $}, normalized to $^{186}$Hg, from experimentally measured isotope shifts 
were compared to MCSM calculations using the extracted $\beta_2$ values (Eq. \ref{eq:beta2_from_Q0}) in combination with 
the nuclear droplet model \cite{Berdichevsky1985} by Eq. \ref{eq:r2} as detailed in \cite{Otten1989,2010Cheal}. 
Panel a) of Fig. \ref{Fig:charge_radius} shows the comparison of MSCM calculations with experimental values.


The ground state of $^{186}$Hg is known from its observed level scheme and from Coulomb-excitation measurements to be only weakly deformed \cite{ensdf,Bree2014}.
This nucleus was used to normalize the experimental \mbox{$\delta <r^2>$}-values from laser-spectroscopy measurements with theoretical calculations. 
The plot in panel a) of Fig. \ref{Fig:charge_radius} highlights $\delta <r^2>_{A-186}$ of the lowest-lying nuclear states with the correct spin 
and parity in a light blue shade if the calculated magnetic moment is similar to the experimentally observed one and in gray when they do not match.
Since even-even nuclei, having $0^+$ ground states, do not have a magnetic moment, 
all even-mass states are indicated in blue on \mbox{Fig.~\ref{Fig:charge_radius}}.
The width of the colored areas corresponds to the spread of MCSM basis-vector deformation parameters on the PES. 
The levels for which both the magnetic moment and deformation agree with experiment are connected by the blue band.\\


An overall agreement for the shape staggering is observed in both the magnitude and location as a function of neutron number. 
In all but $^{181, 185m}$Hg the state corresponding to the electromagnetic moments and charge-radii differences 
observed in experiment is also the lowest state of that given spin and parity.
The difference in energetic ordering for $^{181,185m}$Hg might come from the limit of 24 MCSM basis vectors used in the calculation.
Dedicated calculations with 12 and 16 basis vectors showed indeed that the magnetic and quadrupole moments converged to a large extent
for 24 basis vectors, while the excitation energy of the calculated states could still shift when more basis vectors would be added.


\newpage
\onecolumngrid

\clearpage
\twocolumngrid

For the lightest mercury isotopes with $A \leq 180$, the MCSM correctly describes the trend towards sphericity, magnetic moments and even the energetic ordering of levels.
The unusually rapid and large change of the \mbox{$\delta <r^2>_{A-186}$} value for neutron-deficient mercury isotopes 
with $A<186$ is thus reproduced well in current calculations.  
While the present Hamiltonian is rather standard, the unprecedented size of the configuration space plays an essential role in the calculation outcome.\\

\paragraph*{Shape staggering mechanism}

\begin{figure}[ht]
\begin{center}
\includegraphics[width=1.0\columnwidth]{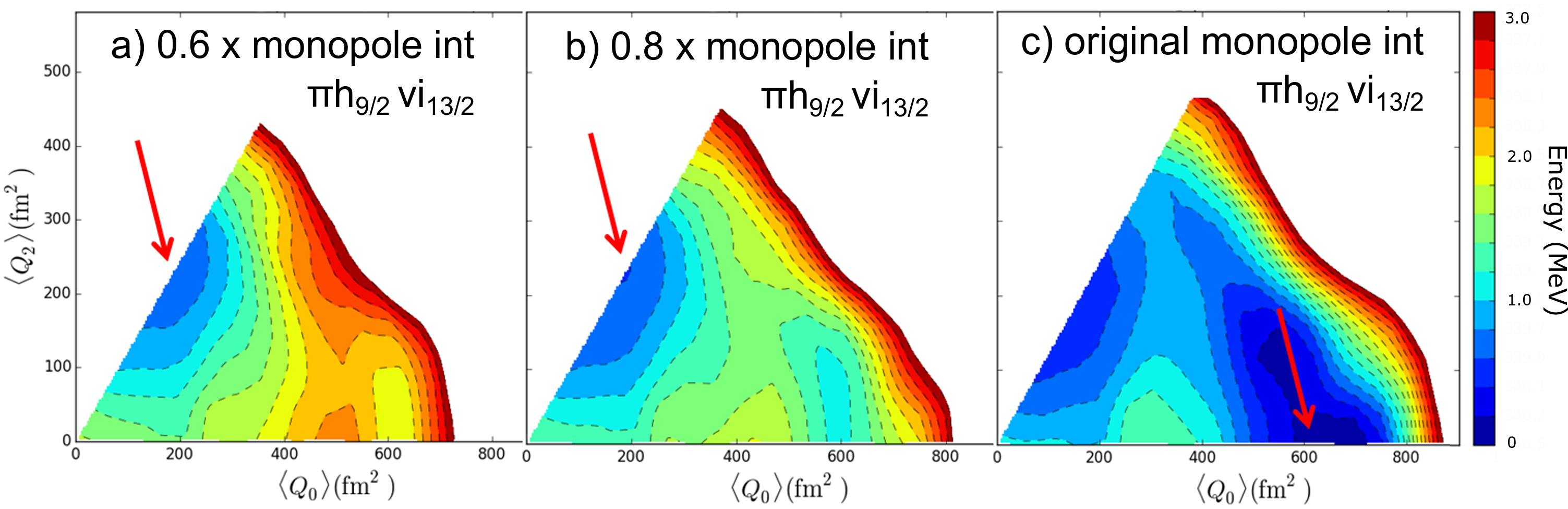}
\caption{Influence of the monopole interaction between $\pi h_{9/2}$ and $\nu i_{13/2}$ on the potential energy surface (PES) of $^{185}$Hg. 
In panels A and B, the interaction is artificially reduced by a factor of 0.6 and 0.8 respectively with respect to the original shown in panel C.
The red arrows indicate the PES minima.}
\label{Fig:Influence_monopole_on_PES}
\end{center}
\end{figure}

The present MCSM calculation enables one to investigate the underlying microscopic factors driving the abrupt change between the near-spherical and deformed states.
Notably, the change in nuclear shape is related to a change in the occupation numbers.
The most important ones are the proton 1$h_{9/2}$ orbital, which is the strongest contributor to protons excitation above $Z=82$, and the neutron 1$i_{13/2}$ orbital, shown in panel c) of Fig.~\ref{Fig:charge_radius}.   
Large and constant values ($\approx 8$) of the occupation number of the neutron 1$i_{13/2}$ orbital are observed for the strongly deformed 1/2$^-$ states of $^{181,183,185}$Hg.
In addition, more than 2 protons are excited above the $Z=82$ closed shell to the 1$h_{9/2}$ orbital.  
For weakly-deformed states, the occupation number of the 1$i_{13/2}$ orbital grows steadily with neutron number, 
while that of the proton 1$h_{9/2}$ orbital remains small as is expected from the usual filling of proton and neutron orbitals.  
The origin of this abrupt change in occupancy numbers as a function of neutron number is found in the monopole component of the $NN$ interaction.
The effect of the monopole interaction between protons in the orbital $j_p$ and neutrons in the orbital $j_n$ can be expressed as 
\begin{equation}
	E_{\text{mon}} = f(j_p, j_n) n_{\pi}(j_p) n_{\nu}(j_n),
\end{equation}
where $f(j_p, j_n)$ is the monopole matrix element, $n_{\pi}(j_p)$ and ($n_{\nu}(j_n)$ stand for the number of protons and neutrons in the specified orbitals, respectively.
The average value of $f(j_p, j_n)$ for different orbitals is about \mbox{-0.2 MeV} in mercury, 
but $f(\pi h_{9/2}, \nu i_{13/2})$ stands out with a strongly-attractive value of -0.35 MeV.  
This is due to the similarity in radial wave functions of the two orbitals 
and due to the effect of the tensor force originating from the attractive $j_>$-$j'_<$ coupling \cite{2016Otsuka_TypeII}.
Once optimal numbers of protons and neutrons are found in these orbitals, 
they produce a large quadrupole deformation, resulting in an increased binding energy due to the proton-neutron quadrupole correlation energy.  
However, these orbitals lie above the Fermi energy and a mechanism is needed to bring such states down in energy.
The strong attractive nature of the monopole interaction provides such a mechanism for $f(\pi h_{9/2}, \nu i_{13/2})$ 
since it is more attractive by 0.15 MeV than the average interaction.
When moving from $^{180}$Hg to $^{181}$Hg, three more protons and six more neutrons in those orbitals, produce an additional monopole binding energy of 2.7 MeV besides the quadrupole contribution.   
The monopole interaction thus shifts the strongly-deformed state down in energy.
The influence this monopole interaction has on the PES is shown in Fig. \ref{Fig:Influence_monopole_on_PES} 
where the strength of this interaction is artificially reduced. 
Here, a shift of the minimum, indicated by the red arrows, is seen from a prolate to oblate shape when the monopole interaction is reduced.

This is a variation of the so-called type-II shell evolution, 
where significant changes in nucleon occupation numbers produce large shifts of effective single-particle energies.
In the present case, the single-particle energies of the neutron 1$i_{13/2}$ orbital and proton 1$h_{9/2}$ orbital are effectively lowered, locating deformed prolate states at lower energy. 
The combined action of the monopole and quadrupole interactions 
thus allows for a near-degenerate coexistence of strongly- and weakly-deformed states in the region of the neutron-deficient mercury istopes.

While this effect is present in all mercury istopes around the neutron mid-shell, 
it is the small additional difference in pairing energy gain between odd-even and even-even nuclei,
which tips the balance in favor of a strongly- or weakly-deformed ground state.

Type-II shell evolution plays an essential role in the quantum phase transition in zirconium isotopes 
and in the shape coexistence in nuclei around $^{68}$Ni \cite{2016QPT_Zr,2014_Ni68}.  
It appears that type-II shell evolution also produces the abrupt odd-even staggering effect in mercury isotopes.
As sucht, this insight links our results to features at different locations of the nuclear chart.
This mechanism, that was qualitatively described in \cite{2011_Heyde_Wood_RevModPhys}, is now put on quantitive grounds.
Moreover, the relevant proton and neutron orbitals are identified. The MCSM calculations can now also be compared to other observables such as energy band structures and transition matrix elements from nuclear spectroscopy, Coulomb excitation and transfer reaction experiments.

\onecolumngrid
\clearpage
\vspace{20pt}
\vspace{40pt}

\twocolumngrid


\section{\label{sec:Conclusion}Conclusion}
We report on an experimental and theoretical study of the neutron-deficient mercury isotopes.
By measuring for the first time the IS and HFS of $^{177-180}$Hg and validating previous measurements along the chain of heavier mercury isotopes, 
the end-point of shape-staggering in the neutron-deficient direction was found to be at $^{180}$Hg.
Not only the $\delta \langle r^2\ \rangle $ of $^{177-180}$Hg, but the magnetic and quadrupole moments of $^{177,179}$Hg 
support the inference of their small deformation.
More advanced descriptions were obtained from the MCSM and DFT calculations.

The DFT models offer an interpretation of the observed difference in charge radii 
in terms of the odd-even mass staggering combined with multiple near-degenerate minima 
of sufficiently different deformation. 
The UNEDF1$^\text{SO}$ with modified pairing strength as well as the SLy5s1 and the SLy5s4 functionals 
provide a radius staggering that is similar to experiment, but predict a staggering pattern that sets in with two neutrons less. 
Reproducing all details of the observed staggering of the radii is not possible at the moment, 
as a quantitative description of the nuclear charge radii is extremely sensitive to many details of the functionals employed. 
For two such aspects, the pairing strength and the surface tension, 
the delicate balance between the minima and its consequence for the nuclear charge radii was shown explicitly.

The largest ever MCSM calculations were performed for this work.
Both the magnetic and quadrupole moments and the radii changes calculated with the MCSM agree with the experimental results to a remarkable extent.
These calculations point to the origin of the shape staggering as coming from a combination of pairing with a variation of type-II shell evolution. 
In the latter, the monopole interaction between the $\nu i_{13/2}$ and $\pi h_{9/2}$ orbitals causes an enhanced occupation of these orbitals which drives the mercury isotopes to deformation. The change in their occupation from one isotope to the other causes their pronounced shape staggering.
This understanding can now be tested in other areas of the nuclear chart, and the MCSM can be used to calculate different experimental observables.


\section{\label{sec:Acknowledgments}Acknowledgments}
We would like to thank to ISOLDE collaboration and technical staff for providing excellent assistance during the experiment.
This project has received funding from the European Union's Horizon 2020 research and innovation programme and the Seventh Framework Programme for Research and Technological Development under grant agreements 262010 (ENSAR), 267194 (COFUND), 289191 (LA$^{3}$NET), 654002 (ERC-2011-AdG-291561-HELIOS).
This work was supported by FWO-Vlaanderen (Belgium), by GOA/2010/010 (BOF KU Leuven) and by the IAP Belgian Science Policy (BriX network P7/12).
This receive support from JSPS and FWO-Vlaanderen under the Japan-Belgium Research Cooperative Program. 
The MCSM calculations were performed on the K computer at RIKEN AICS (hp160211, hp170230).
This work was also supported in part by Priority Issue on Post-K computer 
(Elucidation of the Fundamental Laws and Evolution of the Universe) from MEXT and JICFuS
This project was partially funded by a grant from the UK Science and Technology Facilities Council (STFC): Consolidated Grant ST/L005794/1
The DFT calculations with UNEDF1$^{SO}$ were performed using the DiRAC Data Analytic system 
at the University of Cambridge, operated by the University of Cambridge High Performance Computing Service 
on behalf of the STFC DiRAC HPC Facility (www.dirac.ac.uk). This equipment was funded by BIS National
E-infrastructure capital grant (ST/K001590/1), STFC capital grants ST/H008861/1 and ST/H00887X/1, 
and STFC DiRAC Operations grant ST/K00333X/1. DiRAC is part of the National e-Infrastructure.
DFT calculations with the SLy5sX parameterizations were performed at the Computing Centre of the IN2P3 
and at the Consortium des \'Equipements de Calcul Intensif (C\'ECI), funded by the Fonds de la Recherche 
Scientifique de Belgique (F.R.S.-FNRS) under Grant No. 2.5020.11.
S. S. acknowledges a SB PhD grant from the former Belgian Agency for Innovation by Science and Technology (IWT), 
now incorporated in FWO-Vlaanderen. 
L. P. G. acknowledges FWO-Vlaanderen (Belgium) as an FWO Pegasus Marie Curie Fellow.
J.D. and A.P. acknowledge a partial support from the STFC grant No.\ ST/P003885/1.

\bibliography{Bibliography}

\end{document}